\documentclass[journal]{IEEEtran}
\usepackage{setspace}
\usepackage{tabularx}
\usepackage{array,makecell,longtable,booktabs,multirow}
\usepackage{float}
\usepackage[table]{xcolor}

\definecolor{graymid}{RGB}{242,242,242}
\definecolor{greenlight}{RGB}{238,247,239}
\definecolor{greenmid}{RGB}{217,238,219}
\definecolor{greenstrong}{RGB}{185,220,190}
\usepackage{stfloats}
\usepackage{amsmath,amssymb,mathtools,bm}
\usepackage{graphicx}
\usepackage{booktabs,multirow,array}
\usepackage{xcolor}
\usepackage{cite}
\usepackage{url}
\usepackage{xspace}
\usepackage{stfloats}
\usepackage[hidelinks]{hyperref}
\usepackage[ruled,vlined]{algorithm2e}
\DontPrintSemicolon

\graphicspath{{figures/}}
\newcommand{\vect}[1]{\bm{#1}}

\newcommand{\cK}{\mathcal{K}}
\newcommand{\cA}{\mathcal{A}}
\newcommand{\cM}{\mathcal{M}}
\newcommand{\cC}{\mathcal{C}}
\newcommand{\cF}{\mathcal{F}}
\newcommand{\cG}{\mathcal{G}}

\newcommand{\SAGE}{\textsc{SAGE-HAP}\xspace}

\newtheorem{lemma}{Lemma}
\newtheorem{proposition}{Proposition}
\newtheorem{theorem}{Theorem}

\title{SAGE-HAP: Sustainability-Certified Agentic Communication and Computing at a Solar-Powered High-Altitude Platform}

\author{Haoxiang Luo, Bang Huang, Mohamed-Slim Alouini~\IEEEmembership{Fellow,~IEEE}

\thanks{H. Luo, B. Huang, and M. S. Alouini are with the Computer, Electrical and Mathematical Science and Engineering (CEMSE) Division in King Abdullah University of Science and Technology (KAUST), Thuwal 6900, Makkah Province, Saudi Arabia. (emails: \{haoxiang.luo, bang.huang, slim.alouini\}@kaust.edu.sa).}}

\begin{document}
\maketitle

\begin{abstract}
  The solar-powered High-Altitude Platforms (HAP) can provide persistent communication and computing services beyond terrestrial coverage. However, changing mission priorities, coupled resource demands, and uncertain solar supply complicate sustainable scheduling, while large language model proposals require independent feasibility checks. This paper proposes \SAGE, a Sustainability-certified Agentic Ground-service Engine for a HAP. We formulate an energy-causal stochastic optimization problem incorporating semantic modes, multi-user interference, channel aging, essential platform loads, and battery stress. An agent translates operator intent into typed admission, mode, and grouping candidates, while deterministic convex refinement and an action shield verify resource and reserve feasibility using calibrated solar lower bounds before execution. Agent proposals and deterministic fallbacks share the same certification path. Residual-based forecasting and memory-assisted replanning close the control loop. We establish fixed-candidate convexity, reserve safety under calibrated solar coverage and bounded actuation error, and an online utility bound accounting for candidate-generation error. At nominal load, simulations show 2.38 percentage points higher completion and 5.27\% greater useful-workload energy efficiency than energy-harvesting Lyapunov control. Under stressed load and solar uncertainty, reserve violations occur in 1/50 runs, versus 47/50 without energy certification. A 600-record calibrated inference replay further shows 23.5--36.5\% utility gains over deterministic-only candidates under compound stress.
\end{abstract}


\section{Introduction}
\label{sec:introduction}
\subsection{Background}
\IEEEPARstart{P}{ersistent} connectivity and timely task processing are essential for remote monitoring, maritime operations, and emergency response, yet terrestrial communication and computing infrastructure may be sparse or unavailable. Space--air--ground integrated networks can extend these services beyond terrestrial coverage \cite{luo2026agentic,wang2025toward,luo2025convergence}. A High-Altitude Platform (HAP), operating quasi-stationarily at approximately 20~km altitude, can combine wide-area access with solar generation, battery storage, and onboard computing \cite{kurt2021vision,arum2020review}. The HAP can therefore serve as a regional communication-and-computing anchor: ground devices upload traffic or computation tasks, and the HAP coordinates access, relaying, onboard execution, and result delivery \cite{huang2026joint,huang2026joint1}.

Concentrating these functions on one platform also couples their resource demands. Radio transmission and onboard processing share a finite battery with essential flight and housekeeping loads, while solar input varies over time. Serving more traffic now may consume energy needed to maintain operation after sunset. Service value is also task-dependent. For example, a rescue-control packet and a routine camera update have different priorities and deadlines, while raw data, extracted features, and semantic results incur different communication, computing, and fidelity costs \cite{liang2025knowledge}. The controller must consequently decide which tasks to serve, which representation each task should use, and how much radio and computing capacity can be allocated without compromising future operation.
Accordingly, sustainability means maintaining useful, priority-aware service while preserving a physical battery reserve, respecting long-term energy neutrality, and limiting battery-throughput stress. Semantic mode selection supports this objective by reducing communication and computation where task fidelity permits. These requirements couple immediate service quality to the platform's ability to continue operating \cite{luo2026toward}.

\subsection{Research Motivation}
Existing methods cover only parts of this problem. HAP radio research optimizes coverage, beamforming, or association under communication power budgets \cite{arum2024twotier,javed2025system}. HAP-assisted Mobile Edge Computing (MEC) research addresses offloading and queue management \cite{lakew2021offloading,traspadini2023realtime,ren2023handoff}, while green HAP computing studies investigate solar-powered aerial data centers \cite{abderrahim2023green,abderrahim2024datacenter}. However, a communication power budget or an offloading objective alone does not determine how mission-dependent task value should be balanced against future reserve availability. The remaining need is an online controller that connects changing mission priorities and task representations to interference-aware radio allocation, onboard processing, and reserve preservation.

The time-varying nature of task status further complicates this integration. Deep reinforcement learning (DRL) offers fast decisions, but its policies are optimized for predefined state, action, and reward specifications \cite{luo2024communication,luo2026best}. Akhtarshenas et al.~\cite{akhtarshenas2026ppo} employ Proximal Policy Optimization (PPO) to dynamically position HAP-mounted base stations under stratospheric wind disturbances and maritime user mobility. Furthermore, Li et al.~\cite{li2026efficient} integrate Multi-Agent PPO (MAPPO) with an exact potential game to coordinate task offloading and resource allocation in HAP-assisted Low Earth orbit (LEO) satellite networks, balancing latency, energy consumption, and operational cost. These approaches adapt to network dynamics only within predefined state, action, and reward specifications.
Meanwhile, a standalone Large Language Model (LLM) can interpret a new mission. However, its numerical outputs alone do not establish verifiable link budgets, battery feasibility, or safety margins \cite{luo2026trustworthy}. Specifically, Yan et al.~\cite{yan2026hierarchical} combine HAP-level LLM coordination with Uncrewed Aerial Vehicle (UAV)-level LLM spatial reasoning, a fast double deep Q-network for communication decisions, and deterministic motion decoding. Their framework illustrates how semantic reasoning can be integrated with physical control at different timescales. Its focus on UAV flight and connectivity leaves the joint solar-energy budget, semantic ground-task modes, and per-action battery-reserve certification considered here outside its formulation.

These observations motivate a controller that can translate mission intent into executable service plans and verify their physical consequences before actuation. The difficulty is that intent changes discrete task decisions, those decisions reshape resource consumption, and operating conditions evolve during execution. This gives rise to four coupled challenges.

\begin{itemize}
  \item \emph{C1--Energy-causal communication and computation coupling:}
  Solar-powered HAP communication and station keeping depend on a shared harvested-energy budget \cite{javed2023interdisciplinary}, while finite energy storage couples scheduling decisions across time \cite{huang2013utility}. When onboard computing shares this budget, a currently feasible service action can reduce future operating margins. The challenge is to jointly schedule communication and computation while preserving the flight reserve under uncertain solar supply.

  \item \emph{C2--Intent-to-resource translation:}
    Semantic communication relates information delivery to task objectives \cite{strinati2021semantic}, while semantic extraction couples transmitted data volume with communication and computing costs \cite{cang2023semantic}. For the considered HAP, changing mission priorities and fidelity requirements must be translated into admission decisions and discrete raw, feature, or semantic modes. This enables the feasible and continuous allocation of flight time, launch energy and computing time.

  \item \emph{C3--Reliable actuation from stochastic reasoning:}
  LLM-generated plans may violate constraints, rely on stale observations, or misuse tool outputs \cite{luo2026trustworthy}. Intent adaptation of HAP therefore requires an explicit boundary between proposing an action and authorizing its execution, with independently checked radio, computing, deadline, and reserve feasibility.

  \item \emph{C4--Closed-loop adaptation under compound change:}
   Mobility and environmental disturbances require adaptive HAP control \cite{akhtarshenas2026ppo}, while hierarchical LLM control separates slower reasoning from faster physical execution \cite{yan2026hierarchical}. In the considered system, traffic surges, channel degradation, solar forecast errors, and mission updates may overlap. Thus, the controller must use execution feedback to update predictions and replan while keeping fast scheduling within valid resource limits.

\end{itemize}

\subsection{Our Contributions}
To address these challenges, we propose \SAGE, a \emph{Sustainability-certified agentic Ground-service Engine} for a single solar-powered HAP. \SAGE integrates agentic Artificial Intelligence (AI) for mission interpretation, discrete service planning, and feedback-driven replanning with deterministic tools for resource allocation and battery-reserve certification. Its contributions correspond directly to C1--C4:

\begin{itemize}
  \item \emph{Physical grounding for sustainable HAP autonomy:}
  We develop an energy-causal communication-and-computing model that provides the physical foundation for agentic HAP service planning. A common typed interface represents relay traffic and computation tasks. The model jointly captures multi-user interference, Rician fading, Doppler effects, Channel State Information (CSI), solar harvesting, essential platform loads, and battery stress.

  \item \emph{Intent-aware agentic service orchestration :}
  We formulate a long-term mixed-integer stochastic program that provides a quantitative basis for translating operator intent into HAP service decisions. Task priorities and service ratios guide the agent's admission and semantic-mode proposals, which determine the subsequent continuous resource allocation. Mode selection changes actual input bits, result bits, compute cycles, and fidelity. We also establish the problem's combinatorial hardness.

  \item \emph{Certified execution of agent-generated proposals:}
  We connect agentic planning to physical actuation through a typed candidate interface, a robust convex refiner, and a deterministic action shield. The agent proposes discrete service candidates, while numerical tools determine allocations and verify feasibility before activation. We prove fixed-candidate convexity and reserve safety under the stated solar-coverage and actuation assumptions. We also derive an online utility bound that explicitly accounts for candidate-generation error.

  \item \emph{Closed-loop agentic adaptation and evaluation:}
  We close the HAP control loop through residual-based forecast updates, memory-assisted replanning, and fallback selection, while separating agentic orchestration from fast scheduling. We evaluate the integrated architecture through event-level simulation, a 600-record calibrated inference-trace replay, and Multi-User Multiple-Input Multiple-Output (MU-MIMO) Monte Carlo tests covering changes in mission intent, traffic, channels, and solar supply.
\end{itemize}

The rest of this paper is organized as follows. Section~\ref{sec:related} reviews HAP communication and computing, sustainable control, semantic resource allocation, and agentic network automation. Section~\ref{sec:model} presents the system model. Section~\ref{sec:problem} formulates the optimization problem. Section~\ref{sec:algorithm} develops \SAGE and its analysis. Section~\ref{sec:evaluation} reports the performance evaluation, and Section~\ref{sec:conclusion} concludes the paper.

\section{Related Work}
\label{sec:related}

\subsection{HAP Communication and High-Altitude Edge Computing}

HAP surveys \cite{kurt2021vision,arum2020review} summarize platform architectures, propagation characteristics, coverage design, and radio resource management. Javed and Alouini~\cite{javed2025system} optimized Non-Orthogonal Multiple Access (NOMA)-based HAP communication design. Sarhan et al.~\cite{sarhan2026beamforming} investigated uncertainty-aware massive Multiple-Input Multiple-Output (MIMO) beamforming under imperfect CSI. Akhtarshenas et al.~\cite{akhtarshenas2026ppo} employed PPO to dynamically position HAP-mounted base stations using radio measurements, improving maritime coverage under user mobility and stratospheric wind disturbances.

For HAP-assisted MEC, Lakew et al.~\cite{lakew2021offloading} jointly optimized partial task offloading and communication resources to improve deadline-constrained completion and reduce terminal energy consumption. Traspadini et al.~\cite{traspadini2023realtime} optimized vehicular offloading through queue-based analysis of real-time service probability, whereas Ren et al.~\cite{ren2023handoff} coordinated task splitting and resource allocation to mitigate handoff-induced computing interruptions. Li et al.~\cite{li2026efficient} integrated MAPPO with an exact potential game to coordinate multilayer offloading in HAP-assisted LEO satellite networks, balancing latency, energy consumption, and operational cost.

However, these do not jointly connect multi-user interference and channel aging to semantic task selection and a protected solar-energy reserve. The feasibility of admitted tasks across radio resources, onboard computing, and future battery states therefore remains insufficiently characterized.

\subsection{Solar-Powered and Sustainable HAP Operation}

Javed et al.~\cite{javed2023interdisciplinary} jointly analyzed solar harvesting, station-keeping flight, and communication power, identifying the benefits of coordinated aerodynamic and NOMA design. Abderrahim et al.~\cite{abderrahim2023green} examined the energy-saving opportunities and operational limitations of HAP-hosted data centers. Their subsequent analytical study~\cite{abderrahim2024datacenter} quantified the energy benefits and communication requirements of using aerial data centers to support terrestrial computing. Energy-harvesting communication theory~\cite{ozel2012awgn} characterized transmission under stochastic energy supply. Online scheduling and Lyapunov optimization~\cite{huang2013utility,neely2010stochastic} provided methods for managing energy availability, queues, and long-term utility.

However, energy causality alone does not establish that a prescribed flight reserve is preserved under solar forecast uncertainty. These studies do not jointly connect uncertainty-calibrated reserve protection, battery-throughput stress, and task-dependent service decisions within an online HAP communication-and-computing controller.

\subsection{Semantic-Aware Joint Communication and Computing}

Xie et al.~\cite{xie2021deep} developed learned semantic transmission to preserve message meaning. Strinati and Barbarossa~\cite{strinati2021semantic} established semantic and goal-oriented communication principles that relate information delivery to task accomplishment. Cang et al.~\cite{cang2023semantic} jointly optimized semantic extraction and communication--computation resources to reduce execution delay under energy constraints. Nguyen et al.~\cite{nguyen2026semantic} reviewed semantic communication through theory of mind, generative AI, and joint source--channel coding, identifying opportunities for more effective information representation and delivery.

For satellite--HAP networks, Minani et al.~\cite{minani2024qosem} introduced a semantic-quality metric and mixed-action DRL for joint semantic selection and resource allocation. Their hierarchical multi-agent framework~\cite{minani2025hybrid} further coordinated semantic--bit modes with power, bandwidth, and computing resources across satellite--HAP--ground links. Nkurunziza and Umehara~\cite{nkurunziza2025deepstas} developed DeepSTAS to improve semantic reconstruction through attention-assisted HAP relaying. Their subsequent work~\cite{nkurunziza2026attention} combined attention, and feature transformation to protect HAP-to-ground semantic transmission.

However, they do not connect changing operator intent, mode-dependent service requirements, and future battery reserves under uncertain solar harvesting. A formulation linking raw, semantic modes to data volume, computing demand, deadlines, and per-action reserve certification remains open.

\subsection{Agentic AI for Network Control}

Wang et al.~\cite{wang2024agents} reviewed autonomous-agent architectures incorporating planning, memory, and tool use. For communication networks, Liu et al.~\cite{liu2026tools} organized the tool-engineering lifecycle and demonstrated teacher-guided DRL with a training-time feasibility shield for energy-constrained UAV tool use. Shokouhi and Wong~\cite{shokouhi2026intent} translated operator intent into optimization objectives, user priorities, and coordinated radio-resource decisions in cell-free Open-Radio Access Network (O-RAN).

For aerial networks, Han et al.~\cite{han2025sky} enabled event-driven multi-HAP coordination. Yan et al.~\cite{yan2026hierarchical} combined hierarchical LLM planning, fast reinforcement learning, and deterministic motion decoding for joint flight and connectivity control. Luo et al.~\cite{luo2026agentic} integrated agentic AI and Stackelberg pricing for dependency-aware satellite--UAV offloading. For solar-powered HAPs, Luo et al.~\cite{luo2026background} developed multi-timescale orchestration integrating semantic processing, forecasting, feedback, and pre-execution resource checks.

These studies provide distributed coordination, semantic adaptation, and feasibility-checking mechanisms. A remaining gap is a specified execution interface that binds each accepted semantic-service proposal to robust radio and computing allocations. It also needs to provide battery-reserve certification for each operation under calibrated solar uncertainty.

\section{System Model}
\label{sec:model}

\subsection{Single-HAP Anchor and Task Model}

\begin{figure}[t]
  \centering
  \includegraphics[width=0.47\textwidth]{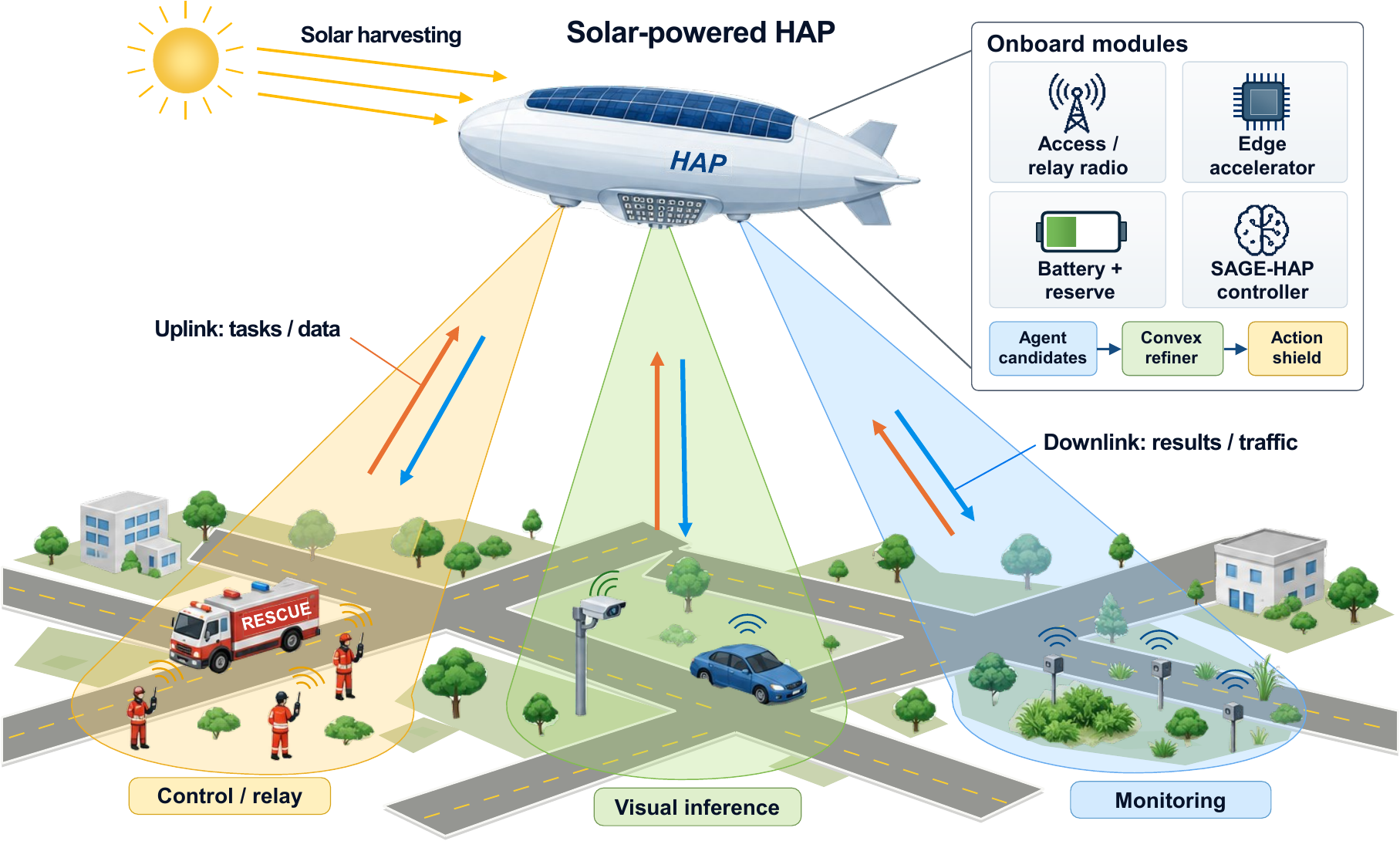}
  \caption{Solar-powered HAP jointly serving relay and computation requests under a protected battery reserve.}
  \label{fig:architecture}
     \vspace{-0.6cm}
\end{figure}

Fig.~\ref{fig:architecture} considers one solar-powered HAP at $\vect q_H=(0,0,H)$, serving ground devices $\cK=\{1,\ldots,K\}$ through an $M$-element array and an edge accelerator. The altitude $H>0$ is quasi-static. Epochs $t=0,1,\ldots$ have duration $\Delta$; Agentic orchestration sets service plans and conventional controllers execute them. Table~\ref{tab:notation} summarizes the principal notation.

\begin{table}[!t]
\centering
\caption{Frequently Used Symbols}
\label{tab:notation}
\renewcommand{\arraystretch}{1.08}
\setlength{\tabcolsep}{4pt}
\footnotesize
\begin{tabular}{p{2cm}p{6.2cm}}
\hline\hline
\textbf{Symbol} & \textbf{Physical meaning} \\
\hline
$\cA_t,\cM_i$ & Eligible requests and request-$i$ service modes. \\
$x_{im,t},y_{i,t}$ & Mode-selection and admission indicators. \\
$\vartheta_{im}$ & Mode-dependent bits, compute cycles, and fidelity. \\
$\cG_{g,t}^s$ & Simultaneous requests in directional group $g$. \\
$\widehat{\vect h}_{k,t}^s,\epsilon_{k,t}^s$ & Predicted channel and uncertainty radius. \\
$\underline g_{k,g}^s,\overline\ell_{kj,g}^s$ & Desired-gain lower bound and leakage upper bound. \\
$e_{i,t}^s,\tau_{g,t}^s$ & Request transmit energy and group airtime. \\
$\tau_{i,t}^c,F^{\max}$ & Compute time and maximum processing rate. \\
$D_{i,t},d_i$ & Scheduled end-to-end delay and deadline. \\
$E_t^h,\underline E_{t+r|t}^h$ & Harvested energy and forecast lower bound. \\
$E_t^{\mathrm{fix}},E_t^{\mathrm{svc}}$ & Essential and planned service energy. \\
$B_t,B^{\min}$, $B^{\max}$ & Battery energy, protected reserve, and capacity. \\
$Q_{\ell,t},Z_t$ & Virtual service and energy deficits. \\
$G_t,V$ & Mission reward and utility--queue weight. \\
$c,\chi_t(c)$ & Discrete plan and its execution certificate. \\
\hline\hline
\end{tabular}
\end{table}

Let $\mathcal L$ be the class set and $\cA_t=\mathcal P_t\cup\cA_t^{\mathrm{new}}$ contain unexpired pending and new requests. Request $i$ and its mode $m\in\cM_i$ are described by
\begin{align}
 \psi_{i,t}&=(k_i,\ell_i,d_i,w_{i,t},a_i^{\min},\cM_i),
 \label{eq:tasktuple}\\
 \vartheta_{im}&=(b_{im}^{u},b_{im}^{d},c_{im},a_{im}).
 \label{eq:modetuple}
\end{align}
Here, $k_i\in\cK$ and $\ell_i\in\mathcal L$ identify the device and class; $d_i>0$ is the deadline from arrival, $w_{i,t}=w_{\ell_i,t}\ge0$ the intent-dependent value, and $a_i^{\min}\in[0,1]$ the fidelity floor. A mode specifies nonnegative uplink/downlink bits $b_{im}^{u},b_{im}^{d}$, HAP cycles $c_{im}$, and fidelity $a_{im}\in[0,1]$, with positive demand in at least one resource. Relay modes have $c_{im}=0$; raw, feature, and semantic-result modes change both data volume and computing demand.

With binary mode variables $x_{im,t}$ and $y_{i,t}=\sum_m x_{im,t}$, admission requires
\begin{equation}
 y_{i,t}\le1,\qquad
 \sum_m x_{im,t}a_{im}\ge a_i^{\min}y_{i,t}.
 \label{eq:modechoice}
\end{equation}
Define $b_{i,t}^s=\sum_m x_{im,t}b_{im}^s$ for $s\in\{u,d\}$ and $C_{i,t}=\sum_m x_{im,t}c_{im}$. Planned class completions are $\widehat S_{\ell,t}=\sum_{i:\ell_i=\ell}y_{i,t}$; measured $S_{\ell,t}$ counts only requests completed within their fidelity and deadline limits. Completed and expired requests leave the pending set.

\subsection{Robust Air-to-Ground Communication}

For horizontal device position $\vect q_{k,t}\in\mathbb R^2$, the slant distance and elevation are
\begin{equation}
 d_{k,t}^{H}=\sqrt{H^2+\|\vect q_{k,t}\|_2^2},\qquad
 \theta_{k,t}=\sin^{-1}(H/d_{k,t}^{H}).
 \label{eq:geometry}
\end{equation}
The large-scale gain $\beta_{k,t}^s$ uses the elevation-dependent propagation model and non-terrestrial parameters in \cite{alhourani2014pathloss,3gpp38811}. At CSI age $T=T_{\mathrm{CSI}}$ seconds, the Rician channel is
\begin{equation}
 \begin{aligned}
 \vect h_{k,t}^s(T)=\sqrt{\frac{\beta_{k,t}^s}{K_k+1}}\big(&
 \sqrt{K_k}\vect a(\theta_{k,t},\varphi_{k,t})
 e^{\mathrm j2\pi f_{D,k}T}\\[-1mm]
 &+\vect z_{k,t}^s(T)\big),
 \end{aligned}
 \label{eq:rician-doppler}
\end{equation}
where $K_k\ge0$ is the Rician factor, $\varphi_{k,t}$ the azimuth, $\mathrm j^2=-1$, and the Uniform Planar Array (UPA) steering vector satisfies $\|\vect a\|_2^2=M$. The Doppler frequency is $f_{D,k}=f_c(\vect v_H-\vect v_k)^{\mathsf T}\vect u_k/c_0$: $f_c$ is the link carrier frequency, $c_0$ the speed of light, $\vect v_H,\vect v_k$ the three-dimensional velocities, and $\vect u_k$ the unit propagation direction under this sign convention. Scattering obeys
\begin{equation}
 \begin{aligned}
 \vect z_{k,t}^s(T)&=\rho_k(T)\vect z_{k,t}^s(0)
       +\sqrt{1-\rho_k(T)^2}\vect\xi_{k,t}^s,\\
 \rho_k(T)&=J_0(2\pi|f_{D,k}|T),
 \end{aligned}
 \label{eq:channel-aging}
\end{equation}
with independent $\vect z_{k,t}^s(0),\vect\xi_{k,t}^s\sim\mathcal{CN}(\vect0,\vect I_M)$ and zeroth-order Bessel function $J_0$. The actuation-time prediction $\widehat{\vect h}_{k,t}^s$ has calibrated radius $\epsilon_{k,t}^s\ge0$:
\begin{equation}
 \|\vect h_{k,t}^s-\widehat{\vect h}_{k,t}^s\|_2
 \le\epsilon_{k,t}^s.
 \label{eq:channel-uncertainty}
\end{equation}
Service guarantees assume this bound holds over the transmission interval.

A candidate assigns each positive-demand request to one group $\cG_{g,t}^s$, indexed by $g_s(i)$, with distinct devices within each group. Stacking rows $\widehat{\vect h}_{k_i,t}^{s\mathsf H}$ gives $\widehat{\vect H}_{g,t}^s$. The link tool forms Regularized Zero-Forcing (RZF) directions \cite{peel2005rzf}:
\begin{equation}
 \widetilde{\vect V}_{g}^{s}
 =\widehat{\vect H}_{g}^{s\mathsf H}
 (\widehat{\vect H}_{g}^{s}\widehat{\vect H}_{g}^{s\mathsf H}
       +\alpha_g\vect I)^{-1},\quad
 \|\vect v_{k,g}^{s}\|_2=1,
 \label{eq:rzf}
\end{equation}
where $\alpha_g>0$ and columns are normalized to obtain downlink precoders or uplink combiners from the corresponding CSI. Superscripts $\mathsf H,\mathsf T$ denote Hermitian transpose and transpose. Suppressing epoch indices, robust desired and leakage gains are
\begin{align}
 \underline g_{k,g}^{s}
 &=\big[|\widehat{\vect h}_k^{s\mathsf H}\vect v_{k,g}^{s}|
          -\epsilon_k^s\big]_+^2,\nonumber\\
 \overline\ell_{kj,g}^{d}
 &=\big(|\widehat{\vect h}_k^{d\mathsf H}\vect v_{j,g}^{d}|
          +\epsilon_k^d\big)^2,\nonumber\\
 \overline\ell_{kj,g}^{u}
 &=\big(|\vect v_{k,g}^{u\mathsf H}\widehat{\vect h}_j^u|
          +\epsilon_j^u\big)^2,
 \label{eq:robustgain}
\end{align}
where $[z]^+=\max\{z,0\}$. Write $a_{kj,g}^{d}=\vect h_k^{d\mathsf H}\vect v_{j,g}^{d}$ and $a_{kj,g}^{u}=\vect v_{k,g}^{u\mathsf H}\vect h_j^u$. For request energy $e_{i,t}^s$ and group airtime $\tau_{g,t}^s$, the physical Signal-to-Interference-plus-Noise Ratio (SINR) and certified service constraints are
\begin{subequations}\label{eq:radio-constraints}
\begin{align}
 &\gamma_{i,g}^{s}
 =\frac{|a_{k_i k_i,g}^{s}|^2e_{i,t}^{s}}
 {N_0W_s\tau_{g,t}^{s}
  +\sum_{j\in\cG_{g,t}^s\setminus\{i\}}
   |a_{k_i k_j,g}^{s}|^2e_{j,t}^{s}},
 \label{eq:sinr}\\
 &\sum_{j\in\cG_{g,t}^s\setminus\{i\}}
 \overline\ell_{k_i k_j,g}^{s}e_{j,t}^{s}
 \le\overline I_{k_i,g}^{s}\tau_{g,t}^{s},
 \label{eq:leakage}\\
 &\phi_{i,g}^s=\tau_{g,t}^{s}W_s\log_2\!\left(1+
 \frac{\underline g_{k_i,g}^{s}e_{i,t}^{s}}
 {(N_0W_s+\overline I_{k_i,g}^{s})\tau_{g,t}^{s}}\right),
 \label{eq:perspective-rate}\\
 &\phi_{i,g_s(i)}^s\ge b_{i,t}^{s},
 \label{eq:bit-service}\\
 &0\le e_{i,t}^{s}\le P_s^{\max}\tau_{g_s(i),t}^{s},
 \label{eq:stream-power}\\
 &\sum_{i\in\cG_{g,t}^d}e_{i,t}^d
 \le P_H^{\max}\tau_{g,t}^d,
 \label{eq:hap-power}\\
 &\sum_{i:k_i=k}e_{i,t}^u\le E_{k,t}^{u,\max},
 \label{eq:user-energy}\\
 &\sum_g\tau_{g,t}^{s}\le\Delta_s,\qquad
 |\cG_{g,t}^{s}|\le M.
 \label{eq:radio-time}
\end{align}
\end{subequations}
Here, $N_0$ is noise spectral density, $W_s$ bandwidth, $\overline I_{k,g}^s$ a tool-fixed interference-power cap, $P_s^{\max}$ a stream-power limit, $P_H^{\max}$ the total HAP downlink-power limit, and $E_{k,t}^{u,\max}$ a verified device-energy budget. Directional airtime budgets are $\Delta_s$. Refinement uses \eqref{eq:leakage}--\eqref{eq:radio-time}; \eqref{eq:sinr} evaluates physical SINR. The concave perspective $\phi_{i,g}^s$ counts bits and equals zero at $(e,\tau)=(0,0)$. Zero-demand directions allocate no energy or task airtime; otherwise $\tau_{i,t}^s=\tau_{g_s(i),t}^s$. Execution preserves the certified groups, beams, and powers $e/\tau$ until completion.

\subsection{Computing and End-to-End Delay}

For positive workload $C_{i,t}$, compute time $\tau_{i,t}^c$ gives processing rate and dynamic energy
\begin{align}
 f_{i,t}&=C_{i,t}/\tau_{i,t}^c,\qquad
 C_{i,t}\le F^{\max}\tau_{i,t}^c,
 \label{eq:frequency}\\
 E_{i,t}^c&=\kappa C_{i,t}^3/(\tau_{i,t}^c)^2,
 \label{eq:compute-energy}\\
 \sum_i\tau_{i,t}^c&\le\Delta_c\le\Delta,\qquad
 \Delta_u+\Delta_d\le\Delta,
 \label{eq:compute-time}
\end{align}
where $F^{\max}$ is the maximum effective processing rate and $\kappa\ge0$ the switched-capacitance coefficient. Set $f_{i,t}=E_{i,t}^c=\tau_{i,t}^c=0$ when $C_{i,t}=0$. Radio operations are serialized; computing may overlap radio activity.

A deterministic compiler creates an acyclic operation graph $(\mathcal O_t(c),\mathcal E_t(c))$ encoding each request's uplink--compute--downlink precedence and the candidate's fixed shared-resource orders. Operation $\nu$ has duration $p_{\nu,t}$ equal to its radio-group or compute time and start $\sigma_{\nu,t}$. Fixed nonnegative edge lags $\delta_{\nu\mu,t}$ cover release and protocol/propagation delays; releases use a zero-time source node. Missing stages are bypassed. Scheduling requires
\begin{equation}
 \begin{aligned}
 &0\le\sigma_{\nu,t},\quad \sigma_{\nu,t}+p_{\nu,t}\le\Delta,\\
 &\sigma_{\mu,t}\ge\sigma_{\nu,t}+p_{\nu,t}+\delta_{\nu\mu,t},
      \quad(\nu,\mu)\in\mathcal E_t(c).
 \end{aligned}
 \label{eq:schedule}
\end{equation}
For final operation $\nu_i^{\mathrm{last}}$, elapsed pre-epoch wait $d_{i,t}^q$, and final overhead $d_{i,t}^0$ not included in the graph,
\begin{equation}
 \begin{aligned}
 D_{i,t}&=d_{i,t}^q+\sigma_{\nu_i^{\mathrm{last}},t}
                  +p_{\nu_i^{\mathrm{last}},t}+d_{i,t}^0,\\
 D_{i,t}&\le d_i,\qquad D_{i,t}-d_{i,t}^q\le\Delta.
 \end{aligned}
 \label{eq:latency}
\end{equation}
These affine constraints include within-epoch resource contention. Unadmitted requests have $D_{i,t}=0$.

\subsection{Solar Energy and Reserve Protection}

For mean irradiance $I_t$, array area $A_{\mathrm{pv}}$, and photovoltaic efficiency $\eta_{\mathrm{pv}}$, harvested and planned service energies are
\begin{align}
 E_t^h&=\eta_{\mathrm{pv}}A_{\mathrm{pv}}I_t\Delta,
 \label{eq:solar}\\
 E_t^{\mathrm{svc}}&=P_{\mathrm{rx}}\sum_g\tau_{g,t}^u\nonumber\\
 &\quad+\sum_i\!\left(e_{i,t}^d+E_{i,t}^c
 +E_c^0\sum_mx_{im,t}\mathbf1_{\{c_{im}>0\}}\right).
 \label{eq:service-energy}
\end{align}
Here, $P_{\mathrm{rx}}$ is receive-circuit power and $E_c^0$ compute activation energy; relay requests incur no activation charge. The known budget $E_t^{\mathrm{fix}}$ covers essential flight, avionics, cooling, and control. Unit amplifier efficiency is assumed; uplink transmit energy is supplied by ground devices.

With battery energy $B_t$, capacity $B^{\max}$, reserve $0<B^{\min}<B^{\max}$, and charging factor $\eta_h\in(0,1]$, the epoch model is
\begin{equation}
 B_{t+1}=\min\!\left\{B^{\max},B_t+\eta_hE_t^h
 -\widetilde E_t^{\mathrm{fix}}-\widetilde E_t^{\mathrm{svc}}\right\}.
 \label{eq:battery}
\end{equation}
Tildes denote metered consumption. Certification uses $E_t^{\mathrm{fix}}+E_t^{\mathrm{svc}}+\Gamma_t$, where $\Gamma_t\ge0$ bounds load, meter, and subepoch accounting error; initial battery telemetry must be correct or conservative. Thermal feasibility uses a characterized upper bound
\begin{equation}
 T_t^{\mathrm{chip}}+\kappa_T(E_t^{\mathrm{chip},0}
 +E_t^{\mathrm{svc}}+\Gamma_t)+\epsilon_T\le T^{\max},
 \label{eq:thermal}
\end{equation}
where $T_t^{\mathrm{chip}}$ is measured temperature, $T^{\max}$ its limit, $\kappa_T>0$ the upper energy-to-temperature coefficient, and $\epsilon_T\ge0$ a model margin. Background chip energy $E_t^{\mathrm{chip},0}$ is already included in $E_t^{\mathrm{fix}}$; cooling receives no credit in this bound.

The forecast supplies point estimates $\widehat E_{t+r|t}^h$ and nonnegative lower bounds for a maximum horizon $L$:
\begin{equation}
 \Pr\!\left(E_{t+r}^h\ge\underline E_{t+r|t}^h,
       \ 0\le r<L\right)\ge1-\delta_h.
 \label{eq:solar-coverage}
\end{equation}
The window risk is $\delta_h\in(0,1)$. Horizonwise conformal residual calibration \cite{angelopoulos2023conformal} requires validity in the operating regime; otherwise a zero lower envelope is used. A convex discharge-stress surrogate \cite{schmalstieg2014aging} is
\begin{equation}
 \begin{aligned}
 \upsilon_t&=[E_t^{\mathrm{fix}}+E_t^{\mathrm{svc}}+
              \Gamma_t-\eta_h\underline E_{t|t}^h]^+,\\
 C_t^{\mathrm{deg}}&=\alpha_b\upsilon_t+\beta_b\upsilon_t^2,
       \qquad\alpha_b,\beta_b\ge0.
 \end{aligned}
 \label{eq:degradation}
\end{equation}
This models discharge stress, not full battery lifetime. Optimization uses the epigraph
\begin{equation}
 \upsilon_t\ge0,\qquad
 E_t^{\mathrm{fix}}+E_t^{\mathrm{svc}}+\Gamma_t
       -\eta_h\underline E_{t|t}^h\le\upsilon_t,
 \label{eq:discharge-epigraph}
\end{equation}
and certification recomputes the positive-part value. The one-step reserve condition is
\begin{equation}
 E_t^{\mathrm{fix}}+E_t^{\mathrm{svc}}
 \le B_t-B^{\min}-\Gamma_t+\eta_h\underline E_{t|t}^h.
 \label{eq:reserve-guard}
\end{equation}
For candidate horizon $L^{(c)}\le L$ and target $R^{(c)}\ge B^{\min}$, define $\overline E_{t|t}^{\mathrm{load}}=E_t^{\mathrm{fix}}+E_t^{\mathrm{svc}}$. Future load bounds $\overline E_{t+r|t}^{\mathrm{load}}$ and margins $\Gamma_{t+r|t}$ are fixed by a specified continuation, such as essential operation without new service; $\Gamma_{t|t}=\Gamma_t$. The certifier computes
\begin{equation}
 \begin{aligned}
 B_{t|t}^{\mathrm{lb}}&=B_t,\\
 B_{t+r+1|t}^{\mathrm{lb}}&=\min\!\big\{B^{\max},B_{t+r|t}^{\mathrm{lb}}
   +\eta_h\underline E_{t+r|t}^h\\
 &\hspace{27mm}-\overline E_{t+r|t}^{\mathrm{load}}
                         -\Gamma_{t+r|t}\big\}.
 \end{aligned}
 \label{eq:reserve-trajectory}
\end{equation}
Requiring every prefix to exceed $R^{(c)}$ prevents reliance on spilled solar energy. Equivalently, refinement uses auxiliary lower energies $\vect b_t=(b_{0|t},\ldots,b_{L^{(c)}|t})$ satisfying
\begin{equation}
 \begin{aligned}
 &b_{0|t}=B_t,\qquad R^{(c)}\le b_{r+1|t}\le B^{\max},\\
 &b_{r+1|t}+\overline E_{t+r|t}^{\mathrm{load}}+\Gamma_{t+r|t}
       \le b_{r|t}+\eta_h\underline E_{t+r|t}^h,
 \end{aligned}
 \label{eq:horizon-feasibility}
\end{equation}
for $0\le r<L^{(c)}$. Only the first-step load depends on the optimized allocation; additional future service requires recertification.

\subsection{Intent, Perception, and Virtual Queues}

Operator intent is parsed into
\begin{equation}
 \iota_t=\{w_{\ell,t},\rho_{\ell,t},d_\ell^{\max},a_\ell^{\min},
 B^{\min},\delta_h,\lambda_D,\lambda_E,\lambda_b\}.
 \label{eq:intent-contract}
\end{equation}
The fields specify class values, target service ratios $\rho_{\ell,t}\in[0,1]$, delay ceilings, fidelity floors, reserve/risk limits, and nonnegative delay, energy, and stress weights. Request contracts obey $d_i\le d_{\ell_i}^{\max}$ and $a_i^{\min}\ge a_{\ell_i}^{\min}$. Invalid updates retain the last valid contract; the agent cannot weaken operator safety limits.

Let $A_{\ell,t}=|\{i\in\cA_t^{\mathrm{new}}:\ell_i=\ell\}|$. Actual completion and energy measurements update virtual deficits
\begin{align}
 Q_{\ell,t+1}&=[Q_{\ell,t}+\rho_{\ell,t}A_{\ell,t}-S_{\ell,t}]^+,
 \label{eq:fairness-queue}\\
 Z_{t+1}&=[Z_t+\widetilde E_t^{\mathrm{fix}}+
          \widetilde E_t^{\mathrm{svc}}-\eta_hE_t^h]^+.
 \label{eq:energy-queue}
\end{align}
Mean-rate stability enforces average service and energy balance; physical battery safety additionally requires the reserve constraints. The verified observation is
\begin{equation}
 \begin{split}
 o_t=\{&\cA_t,\widehat{\vect H}_t,
 \boldsymbol\epsilon_t^{\mathrm{CSI}},\vect n_t,B_t,I_t,\\
 &\widehat{\vect E}_{t:t+L-1}^h,T_t^{\mathrm{chip}},
 \vect Q_t,Z_t,\iota_t\},
 \end{split}
 \label{eq:observation}
\end{equation}
where $\widehat{\vect H}_t$ collects directional CSI, $\boldsymbol\epsilon_t^{\mathrm{CSI}}$ its radii, $\vect n_t$ physical queue occupancies, and $\vect Q_t=(Q_{\ell,t})_{\ell\in\mathcal L}$. Associated metadata bind geometry, velocity, arrival times, budgets, timestamps, and calibration versions.

\section{Problem Formulation}
\label{sec:problem}

The per-epoch certified mission reward is
\begin{align}
 G_t={}&\sum_{i\in\cA_t}\sum_{m\in\cM_i}
 x_{im,t}w_{i,t}a_{im}
 -\lambda_D\sum_i y_{i,t}\frac{D_{i,t}}{d_i}
 \nonumber\\
 &-\lambda_E\left(E_t^{\mathrm{svc}}
                 +\omega_u\sum_i e_{i,t}^{u}\right)
 -\lambda_b C_t^{\mathrm{deg}},
 \label{eq:reward}
\end{align}
where $\omega_u\ge0$ is the relative weight of ground-device energy. This reward values service supported by a feasible certificate and prices its reserved resources. Executed completion, energy, and delay are recorded separately when a plan is interrupted.

Let $\Pi$ denote a causal sequence of discrete plans, radio energies, operation durations, start times, and auxiliary discharge/reserve variables. Each plan fixes admission, modes, grouping, resource order, reserve target, and continuation assumptions. The sustainable service problem is
\begin{subequations}\label{prob:p0}
\begin{align}
 &\mathbf{P0}:\quad
 \max_{\Pi}\quad
 \liminf_{T\to\infty}\frac1T
 \sum_{t=0}^{T-1}\mathbb E[G_t],
 \label{p0:obj}\\
 &\mathrm{s.t.}\quad
 \eqref{eq:modechoice},\
 \eqref{eq:radio-constraints},\
 \eqref{eq:frequency},\
 \eqref{eq:compute-time},
 \label{p0:instant}\\
 &\eqref{eq:schedule},\
 \eqref{eq:latency},\
 \eqref{eq:thermal},\
 \eqref{eq:discharge-epigraph},
 \nonumber\\
 &B_t\in[B^{\min},B^{\max}],\
 \eqref{eq:battery},\
 \eqref{eq:reserve-guard},\
 \eqref{eq:horizon-feasibility},
 \label{p0:battery}\\
 &\limsup_{T\to\infty}\frac1T
   \sum_{t=0}^{T-1}
   \mathbb E[\rho_{\ell,t}A_{\ell,t}-S_{\ell,t}]
   \le0,\quad\ell\in\mathcal L,
 \label{p0:fairness}\\
 &\limsup_{T\to\infty}\frac1T
   \sum_{t=0}^{T-1}\mathbb E[
   \widetilde E_t^{\mathrm{fix}}+\widetilde E_t^{\mathrm{svc}}
   -\eta_hE_t^h]\le0,
 \label{p0:neutrality}\\
 &x_{im,t}\in\{0,1\},\quad e_{i,t}^s\ge0,\nonumber\\
 &\tau_{g,t}^s,\tau_{i,t}^c,
 \sigma_{\nu,t},\upsilon_t\ge0.
 \label{p0:domain}
\end{align}
\end{subequations}
Constraint~\eqref{p0:instant} enforces per-epoch admission, fidelity, robust transmission, computing, scheduling, deadline, and thermal feasibility; the discharge epigraph provides a convex representation of the discharge-energy proxy. 
Constraint~\eqref{p0:battery} enforces finite-capacity battery dynamics and protects the flight reserve for the current action and its specified continuation under valid solar and load bounds.
Constraint~\eqref{p0:fairness} requires long-term completed service to cover target-weighted arrivals for each class; constant $\rho_{\ell,t}=\rho_\ell$ and a positive average arrival rate recover the usual service-ratio requirement.
Constraint~\eqref{p0:neutrality} imposes long-term energy neutrality: average essential and service consumption cannot exceed usable harvested energy.
Constraint~\eqref{p0:domain} specifies binary admission/mode choices and nonnegative transmit energies, radio/compute durations, operation start times, and discharge proxies. Instantaneous constraints apply to all active requests, groups, and operations. Planning uses $\widehat S_{\ell,t}$, whereas \eqref{eq:fairness-queue} is updated with measured completions $S_{\ell,t}$; the online analysis states the required service-realization assumptions.

\textbf{P0} couples discrete admission, semantic selection, grouping, and order with continuous resource allocation. Current actions also change the battery and pending-request states, while arrivals, CSI, solar supply, and operator intent are only causally observed. Feasible service is considered on states where essential platform operation itself can be supplied; an empty-service proposal cannot overcome an unavoidable essential-energy deficit.

\begin{proposition}[Combinatorial hardness]
Even a one-epoch restriction of \textbf{P0}, with one mode per request, fixed channels, zero queueing delay, and no discharge penalty, is NP-hard.
\label{prop:hardness}
\end{proposition}
\begin{IEEEproof}
Set communication demands to 0, make energy and thermal limits nonbinding, and give all requests loose deadlines. Assign request $i$ compute demand $c_i>0$ and value $v_i=w_i a_i>0$. With maximum processing rate $F^{\max}$ and compute budget $\Delta_c$, a selected subset is schedulable exactly when $\sum_i x_i c_i\le C$, where $C=F^{\max}\Delta_c$ is the epoch cycle budget: necessity follows from \eqref{eq:frequency}, and sufficiency follows by sequential execution at $F^{\max}$. Setting the reward penalties to 0 yields $\max\sum_i x_i v_i$ subject to this constraint and $x_i\in\{0,1\}$, which is the 0--1 knapsack problem.
\end{IEEEproof}

This structure motivates a division of labor. agentic reasoning proposes a limited set of discrete, intent-consistent plans. Deterministic tools establish the physical coefficients and operation order, after which continuous refinement has the convex structure developed in Section~\ref{sec:algorithm}.

\section{SAGE-HAP: Certified Agentic Control}
\label{sec:algorithm}

\subsection{Typed Planning and Control Separation}

\begin{figure*}[t]
  \centering
  \includegraphics[width=0.98\textwidth]{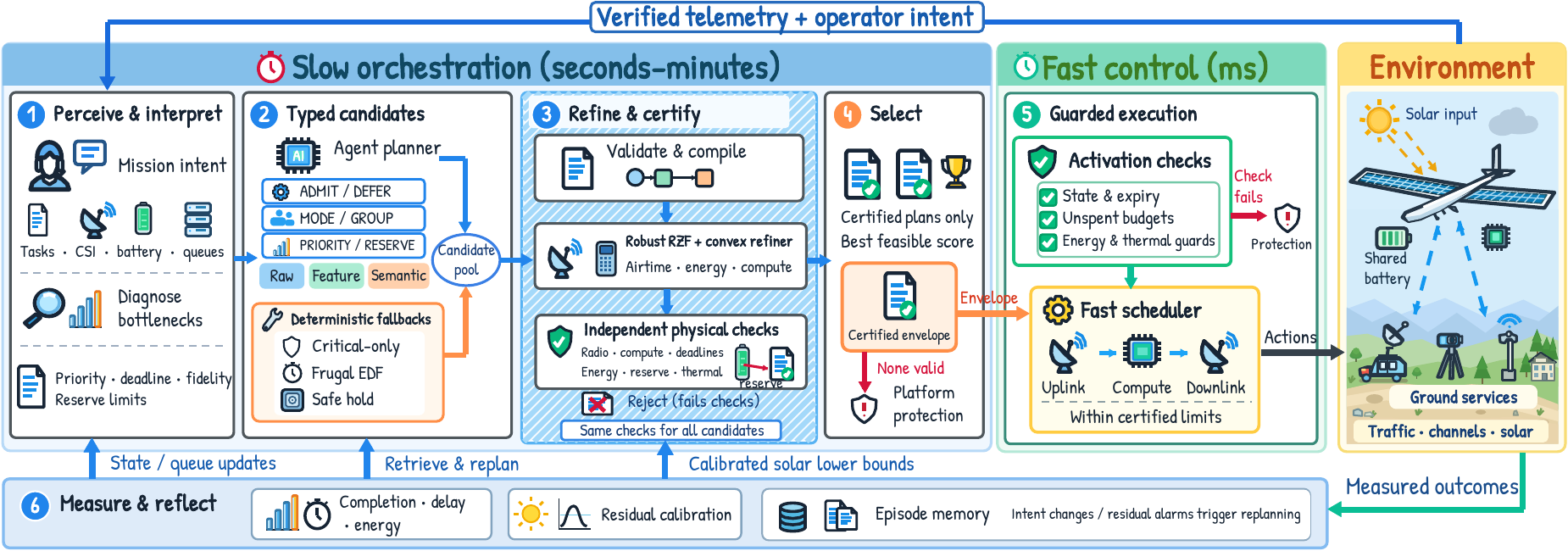}
  \caption{Closed-loop architecture of SAGE-HAP. Slow orchestration generates intent-aware candidates and deterministic fallbacks, which share the same refinement and certification path. Fast control executes selected plans within certified resource limits. Measured outcomes update system states, forecast calibration, and episode memory, closing the loop through feedback-driven replanning.}
  \label{fig:loop}
    \vspace{-0.5cm}
\end{figure*}

In Fig.~\ref{fig:loop}, agentic AI interprets intent, diagnoses bottlenecks, retrieves experience, and proposes discrete plans. Deterministic tools alone allocate resources and authorize actuation. Conventional controllers execute on the millisecond timescale; orchestration operates over epochs of seconds to minutes, while calibration and long-horizon planning refresh over minutes. Algorithms~\ref{alg:sage_main} and~\ref{alg:sage_certify} summarize this separation.

A candidate is
\begin{equation}
 c=(\{x_{im}^{(c)}\},\{\cG_g^{s(c)}\},
        \pi^{(c)},L^{(c)},R^{(c)},\zeta^{(c)}),
 \label{eq:candidate}
\end{equation}
where $\pi^{(c)}$ is class priority with earliest-deadline-first tie breaking, $1\le L^{(c)}\le L$, $B^{\min}\le R^{(c)}\le B^{\max}$, and $\zeta^{(c)}$ binds tool versions and continuation assumptions. The compiler expands class-level \texttt{ADMIT/DEFER}, \texttt{MODE}, \texttt{GROUP}, \texttt{PRIORITY}, and \texttt{RESERVE} primitives into request identifiers and complete resource orders. The agent cannot issue powers, beam vectors, frequencies, or device commands.

The proposal set is $\cC_t=\operatorname{Unique}(\cC_t^{\mathrm A}\cup\cC_t^{\mathrm F})$, combining agent proposals with critical-only, energy-frugal earliest-deadline-first, and safe-hold fallbacks. The same certifier checks every proposal; safe hold still requires essential energy. Let $\cC_t^{\mathrm{ok}}$ be the currently certified subset and $\mathcal M_t$ the episode archive.

\begin{algorithm}[!t]
\caption{Closed-Loop Service Control by \SAGE}
\label{alg:sage_main}
\scriptsize
\SetAlgoSkip{smallskip}
\DontPrintSemicolon
\SetKwInOut{Input}{Input}
\SetKwInOut{Output}{Output}
\Input{Valid intent, physical limits, calibrated tools, and archive $\mathcal M_0$}
\Output{Certified service actions and updated queues/archive}
Initialize $\vect Q_0=\vect0$ and $Z_0=0$\;
\For{orchestration epoch $t=0,1,\ldots$}{
    Verify telemetry and intent; construct $o_t$ by \eqref{eq:observation}\;
    Retrieve $\mathcal M_t$ and generate/deduplicate agent and fallback plans $\cC_t$ within the planning deadline\;
    Set $\cC_t^{\mathrm{ok}}\leftarrow\varnothing$\;
    \ForEach{$c\in\cC_t$, checking fallbacks first}{
        Call Algorithm~\ref{alg:sage_certify}; retain accepted score/certificate pairs in $\cC_t^{\mathrm{ok}}$\;
    }
    Revalidate state, expiry, and unspent budgets; remove stale certificates\;
    \eIf{$\cC_t^{\mathrm{ok}}\ne\varnothing$}{
        Select by \eqref{eq:candidate-selection} and execute under fast energy/thermal guards\;
    }{
        Invoke platform protection; continue an earlier envelope only if still valid\;
    }
    Measure completion/energy; update $B_{t+1},\vect Q_{t+1},Z_{t+1}$ by \eqref{eq:battery}, \eqref{eq:fairness-queue}, and \eqref{eq:energy-queue}\;
    Compute \eqref{eq:residual}, update calibration data and $\mathcal M_{t+1}$; trigger replanning on alarms or intent changes\;
}
\end{algorithm}

\subsection{Convex Refinement and Action Certification}

For fixed $c$, the tools fix RZF directions, leakage caps, operation orders, and future load bounds. Let $\cF_t(c)$ contain the radio, processing, scheduling, thermal, discharge-epigraph, and horizon constraints in Section~\ref{sec:model}; the horizon constraints imply \eqref{eq:reserve-guard}. With energy, duration, and start-time vectors $\vect e,\vect\tau,\vect\sigma$, the refiner solves
\begin{align}
 \Phi_t(c)=\min_{\vect e,\vect\tau,\vect\sigma,\upsilon_t,\vect b_t}
 \quad&V\lambda_D\sum_i y_{i,t}^{(c)}D_{i,t}/d_i\nonumber\\
 &+(Z_t+V\lambda_E)E_t^{\mathrm{svc}}
       +V\lambda_E\omega_u\sum_i e_{i,t}^u\nonumber\\
 &+V\lambda_b(\alpha_b\upsilon_t+\beta_b\upsilon_t^2)\nonumber\\
 &-V\sum_{i,m}x_{im,t}^{(c)}w_{i,t}a_{im}
       -\sum_\ell Q_{\ell,t}\widehat S_{\ell,t}^{(c)}\nonumber\\
 \mathrm{s.t.}\quad&
 (\vect e,\vect\tau,\vect\sigma,\upsilon_t,\vect b_t)\in\cF_t(c).
 \label{prob:p1}
\end{align}
Here, $V>0$ weights reward against virtual deficits; action-independent arrival, essential-load, and harvesting terms are omitted.

\begin{lemma}[Fixed-candidate convexity]
With valid fixed tool bounds, \eqref{prob:p1} is convex.
\label{lem:convex}
\end{lemma}
\begin{IEEEproof}
The bit perspective is concave, whereas $\kappa C_{i,t}^3/(\tau_{i,t}^c)^2$ and its zero-workload extension are convex. Hence bit-service superlevels, thermal/energy upper bounds, and the discharge epigraph are convex. Scheduling and resource constraints are affine; horizon inequalities contain a convex first-step load and constant future loads. The objective combines these convex terms with nonnegative coefficients \cite{boyd2004convex}.
\end{IEEEproof}

The solver returns a feasible allocation with recomputed score $\widehat\Phi_t(c)$ and certified objective gap $0\le\widehat\Phi_t(c)-\Phi_t(c)\le\epsilon_s$. Independent physical checks produce
\begin{equation}
 \begin{split}
 \chi_t(c)=\big(&c,\vect e^\star,\vect\tau^\star,\vect\sigma^\star,
 \upsilon_t^\star,h(\vect V),r_{\max},\\
 &\vect B_{t+1:t+L^{(c)}|t}^{\mathrm{lb}},h(o_t),
 \mathfrak t_t^{\mathrm{iss}},\mathfrak t_t^{\mathrm{exp}}\big).
 \end{split}
 \label{eq:certificate}
\end{equation}
Here, $h$ is a cryptographic hash, $\vect V$ collects beams/combiners, and $\mathfrak t^{\mathrm{iss}},\mathfrak t^{\mathrm{exp}}$ are controller-clock issue/expiry times. The largest normalized constraint residual $r_{\max}$ uses fixed positive constraint scales; its tolerance $\varepsilon_{\mathrm{cert}}\ge0$ is absorbed by inward physical margins. State and tool bindings must remain valid until activation. The selected plan is
\begin{equation}
 c_t\in\arg\min_{c\in\cC_t^{\mathrm{ok}}}\widehat\Phi_t(c).
 \label{eq:candidate-selection}
\end{equation}

\begin{algorithm}[!t]
\caption{Deterministic Refinement and Certification}
\label{alg:sage_certify}
\scriptsize
\SetAlgoSkip{smallskip}
\DontPrintSemicolon
\SetKwInOut{Input}{Input}
\SetKwInOut{Output}{Output}
\Input{Candidate $c$, verified $o_t$, tool bounds, and tolerances $\epsilon_s,\varepsilon_{\mathrm{cert}}$}
\Output{$(\widehat\Phi_t(c),\chi_t(c))$ or \texttt{REJECT}}
Validate the grammar, contract, and tool assumptions; compile the operation graph\;
\If{validation fails or the graph is cyclic}{\Return \texttt{REJECT}\;}
Form normalized RZF directions by \eqref{eq:rzf} and robust gains by \eqref{eq:robustgain}; fix leakage caps and continuation bounds\;
Solve \eqref{prob:p1} within the planning deadline\;
\If{no feasible point with objective gap $\le\epsilon_s$ is available}{\Return \texttt{REJECT}\;}
Independently recompute service, delay, energy, stress, $\widehat\Phi_t(c)$, and $r_{\max}$\;
Recompute all reserve prefixes by \eqref{eq:reserve-trajectory}; verify state binding and expiry\;
\If{$r_{\max}>\varepsilon_{\mathrm{cert}}$ or any physical/freshness check fails}{\Return \texttt{REJECT}\;}
Issue $\chi_t(c)$ by \eqref{eq:certificate}\;
\Return $(\widehat\Phi_t(c),\chi_t(c))$\;
\end{algorithm}

\begin{proposition}[Reserve safety]
With valid telemetry and load-error bounds, \eqref{eq:reserve-guard} and $E_t^h\ge\underline E_{t|t}^h$ ensure $B_{t+1}\ge B^{\min}$. For a fixed forecast window, following its certified continuation preserves the epoch-boundary reserve with probability at least $1-\delta_h$ under \eqref{eq:solar-coverage} and its load bounds.
\label{prop:safety}
\end{proposition}
\begin{IEEEproof}
Substitution into \eqref{eq:battery} gives
$B_{t+1}\ge\min\{B^{\max},B_t+\eta_h\underline E_{t|t}^h-E_t^{\mathrm{fix}}-E_t^{\mathrm{svc}}-\Gamma_t\}\ge B^{\min}$.
Monotonicity and induction on \eqref{eq:reserve-trajectory} yield $B_{t+r}\ge B_{t+r|t}^{\mathrm{lb}}\ge R^{(c)}$ on the joint coverage event.
\end{IEEEproof}

For rolling windows, a finite sequence has solar-bound failure probability at most $\sum_t\delta_{h,t}$; a mission-wide risk budget must allocate this sum.

\subsection{Guarded Execution and Reflection}

The certificate fixes admission, modes, radio profiles, compute allocations, start/completion windows, and energy ledgers. The shield rechecks state and expiry before activation and prevents budget reuse. Fast battery/thermal guards check measured available energy before each hardware interval, without spending unharvested solar energy. A failed check suspends optional service and invokes platform protection, assuming essential flight/control loads remain physically supportable; interrupted requests do not count as completed.

For measured and predicted outcomes $y_t^j,\widehat y_t^j$, normalized residuals are
\begin{equation}
 r_t^j=\frac{y_t^j-\widehat y_t^j}
 {\max\{|\widehat y_t^j|,\epsilon_{y,j}\}},
 \qquad j\in\{h,g,D,E,S\}.
 \label{eq:residual}
\end{equation}
The components denote solar energy, link gain, delay, service energy, and completion; $\epsilon_{y,j}>0$ has matching units, and vectors are processed componentwise. These outcome symbols differ from admission $y_{i,t}$. The archive stores
\begin{equation}
 \mathfrak m_t=(h(\mathrm{context}),c_t,
 \widehat{\vect y}_t,\vect y_t,\vect r_t,\mathrm{feedback}).
 \label{eq:memory}
\end{equation}
No activation produces a null-action record. Residuals update calibration data; new profile bounds require validation. Changed intent, residual alarms, bottleneck mismatch, or rejection of all agent plans triggers replanning. Retrieval uses task mix, solar regime, battery band, and bottleneck, while measured history and operator limits remain binding.

\subsection{Online Performance and Complexity}

The performance bound assumes a fixed regime with realized certificates: $S_{\ell,t}=\widehat S_{\ell,t}$ and actual total load equal to $E_t^{\mathrm{fix}}+E_t^{\mathrm{svc}}$. Let $\mathcal D_t^{\mathrm{feas}}$ contain all feasible discrete plans under the same tools, schedule, and continuation model. For the nonempty surviving set $\cC_t^{\mathrm{ok}}\subseteq\mathcal D_t^{\mathrm{feas}}$, define
\begin{equation}
 \Delta_t^A=\min_{c\in\cC_t^{\mathrm{ok}}}\Phi_t(c)
 -\min_{c\in\mathcal D_t^{\mathrm{feas}}}\Phi_t(c)\ge0.
 \label{eq:candidate-gap}
\end{equation}
Thus, missed proposals and revalidation losses enter $\Delta_t^A$, while $\epsilon_s$ captures numerical error. Let $\mathcal H_t$ denote pre-observation history, including controlled battery and pending states, and set $u_{\ell,t}=\rho_{\ell,t}A_{\ell,t}-S_{\ell,t}$ and $u_{E,t}=\widetilde E_t^{\mathrm{fix}}+\widetilde E_t^{\mathrm{svc}}-\eta_h E_t^h$. A causal comparator feasible at these same states must satisfy
\begin{equation}
 \begin{aligned}
 \mathbb E[u_{\ell,t}^\star\mid\mathcal H_t]&\le0,
 &\mathbb E[u_{E,t}^\star\mid\mathcal H_t]&\le0,\\
 \mathbb E[G_t^\star\mid\mathcal H_t]&\ge G^\star.
 \end{aligned}
 \label{eq:comparison-condition}
\end{equation}
The benchmark $G^\star$ is restricted to this conditional comparison class.

\begin{theorem}[Candidate-aware online bound]
Assume bounded queue increments, $G_{\min}\le G_t\le G_{\max}$, finite initial queues, the realization/comparison conditions above, and $\mathbb E[\Delta_t^A\mid\mathcal H_t]\le D_A<\infty$. Then
\begin{equation}
 \overline G_{\mathrm{SAGE}}\ge
 G^\star-\frac{C_0+\overline\Delta^A+\epsilon_s}{V},
 \label{eq:utility-bound}
\end{equation}
where $\overline G_{\mathrm{SAGE}}=\liminf_{T\to\infty}T^{-1}\sum_{t<T}\mathbb E[G_t]$, $\overline\Delta^A=\limsup_{T\to\infty}T^{-1}\sum_{t<T}\mathbb E[\Delta_t^A]$, and $C_0$ uniformly bounds $\tfrac12\mathbb E[u_{E,t}^2+\sum_\ell u_{\ell,t}^2\mid\mathcal H_t]$. If a feasible causal policy additionally has every conditional queue increment at most $-\epsilon_0<0$ in fixed queue units, then
\begin{equation}
 \overline{\mathcal Q}\le
 \frac{C_0+D_A+\epsilon_s+V(G_{\max}-G_{\min})}{\epsilon_0},
 \label{eq:queue-bound}
\end{equation}
where $\overline{\mathcal Q}=\limsup_{T\to\infty}T^{-1}\sum_{t<T}\mathbb E[Z_t+\sum_\ell Q_{\ell,t}]$. This is $O(V)$ when $D_A=O(V)$ and the other constants are fixed.
\label{thm:online}
\end{theorem}
\begin{IEEEproof}
For $\mathcal V_t=\tfrac12(Z_t^2+\sum_\ell Q_{\ell,t}^2)$, the queue updates give
\begin{align*}
 &\mathbb E[\mathcal V_{t+1}-\mathcal V_t-VG_t\mid\mathcal H_t]\\
 &\quad\le C_0+\mathbb E[Z_tu_{E,t}
       +\textstyle\sum_\ell Q_{\ell,t}u_{\ell,t}-VG_t\mid\mathcal H_t].
\end{align*}
Refinement minimizes the action-dependent right side within $\Delta_t^A+\epsilon_s$. Comparing with \eqref{eq:comparison-condition} and telescoping proves \eqref{eq:utility-bound}. The slack comparator instead contributes $-\epsilon_0(Z_t+\sum_\ell Q_{\ell,t})$; using the reward range and telescoping yields \eqref{eq:queue-bound}. Bounded increments and strong stability imply mean-rate stability and hence \eqref{p0:fairness}--\eqref{p0:neutrality} \cite{neely2010stochastic}.
\end{IEEEproof}

For $n_c$ admitted requests, refinement has $O(n_c+L^{(c)})$ variables. With a comparable-size cone representation, dense interior-point scaling is $O((n_c+L^{(c)})^{3.5}\log(1/\epsilon_{\mathrm{IP}}))$, where $\epsilon_{\mathrm{IP}}>0$ is a numerical tolerance distinct from objective error $\epsilon_s$ \cite{boyd2004convex}. RZF costs $O(Mn_g^2+n_g^3)$ for $n_g=|\cG_g^s|$ streams; serial solve costs add over $\cC_t$. Fast scheduling sorts $n=|\cA_t|$ requests in $O(n\log n)$ time.

\section{Performance Evaluation}
\label{sec:evaluation}

\subsection{Simulation Setup}

We implement a 120-epoch event-level simulation in Python~3.12; one epoch is one minute of orchestration, and the fast scheduler executes typed tasks inside each certified envelope. All experiments
are conducted on a PC equipped with an Intel Core i9-14900HX CPU
(2.2~GHz) and an NVIDIA GeForce RTX 4070 GPU with 8~GB of memory. Control results use 50 seeds and 95\% confidence intervals. For each spatial group, the simulator rebuilds RZF directions from aged CSI, evaluates \eqref{eq:sinr}, and admits service only through \eqref{eq:perspective-rate}; hence, no controller uses an interference-free aggregate gain. Candidate-path tests use 600 observable trace records: 100 runs for each Qwen3-8B and GPT-5.4-mini  profile under nominal, intent-shift, and compound-stress states.

The HAP altitude is 20~km, and the ground-service radius is 10~km. Access uses 28~GHz, four 100-MHz resource groups, a 64-element $8\times8$ UPA, 40~W total downlink Radio Frequency (RF) power, Rician $K=10$~dB, and 2-ms CSI age. Ground speed varies from 0 to 30~m/s. The HAP has an 80~kWh battery, an initial 31\% state of charge (SoC), a 25\% reserve, 15~kW initial solar input declining toward sunset, 7.5~kW fixed station-keeping/avionics load, 0.2--1.5~kW communication-payload power, and 0.3--2.5~kW compute-payload power. Its effective compute capacity is normalized to a 550-TOPS accelerator. Table~\ref{tab:tasks} gives task modes.
The nominal class mix is 27\% control/relay, 34\% visual inference, and 39\% monitoring. At epoch 42, the operator changes intent to prioritize rescue control and visual detection. A compound stress disturbance begins at epoch 48: offered load increases, access capacity is reduced, and the load peaks around epoch 67 while solar input falls. Forecast Root-Mean-Square Error (RMSE) is 1.2~kW unless otherwise stated.

\begin{table}[t]
\caption{Representative Task and Semantic-Mode Parameters}
\label{tab:tasks}
\centering
\renewcommand{\arraystretch}{1.12}
\setlength{\tabcolsep}{3.5pt}
\scriptsize
\begin{tabular}{llrrrr}
\hline\hline
\textbf{Class} & \textbf{Mode} & $b^u$ & $c$ & $a$ & $d^{\max}$ \\
 & & (Mbit) & (Gcycle) & & (ms) \\
\midrule
Control/relay & lossless & 0.04 & 0 & 1.00 & 30 \\
\midrule
Vision & raw & 12.0 & 8.0 & 1.00 & 300 \\
 & feature & 3.6 & 5.6 & 0.95 & 300 \\
 & semantic & 1.14 & 3.04 & 0.87 & 300 \\
\midrule
Monitoring & raw & 2.0 & 1.0 & 1.00 & 3000 \\
 & feature & 0.65 & 0.71 & 0.96 & 3000 \\
 & semantic & 0.22 & 0.39 & 0.89 & 3000 \\
\hline\hline
\end{tabular}
\end{table}

We compare four schemes.
\emph{1) HAP-EDF} admits raw-mode work in Earliest-Deadline-First (EDF) order until radio or compute capacity is reached, following conventional queue-oriented HAP service;
\emph{2) MAPPO} \cite{li2026efficient} adapts centralized training and decentralized execution to HAP service scheduling and resource allocation. Its policy responds to observed task, channel, and battery states, while reward weights and policy parameters remain fixed during evaluation;
\emph{3) EH-Lyapunov} \cite{wang2026lyapunov} applies Energy-Harvesting (EH) drift-plus-penalty control with fairness debt and fixed feature modes, following the online-control principle;
\emph{4) Robust-MPC} \cite{zhang2025maintaining} uses a receding-horizon point solar forecast and robust channel factors, but fixed task values and no agentic reflection. It is based on Model Predictive Control (MPC).

Useful-workload energy efficiency is the fidelity-weighted input-equivalent work completed per joule of HAP payload energy. Mission utility uses the active intent weights. These metrics are reported together with completion, deadline-normalized delay, and final battery State of Charge (SoC) so that semantic compression cannot appear beneficial merely by discarding task value.

\subsection{End-to-End Agent Candidate Trace Replay}

\begin{figure*}[t]
  \centering
  \includegraphics[width=0.95\textwidth]{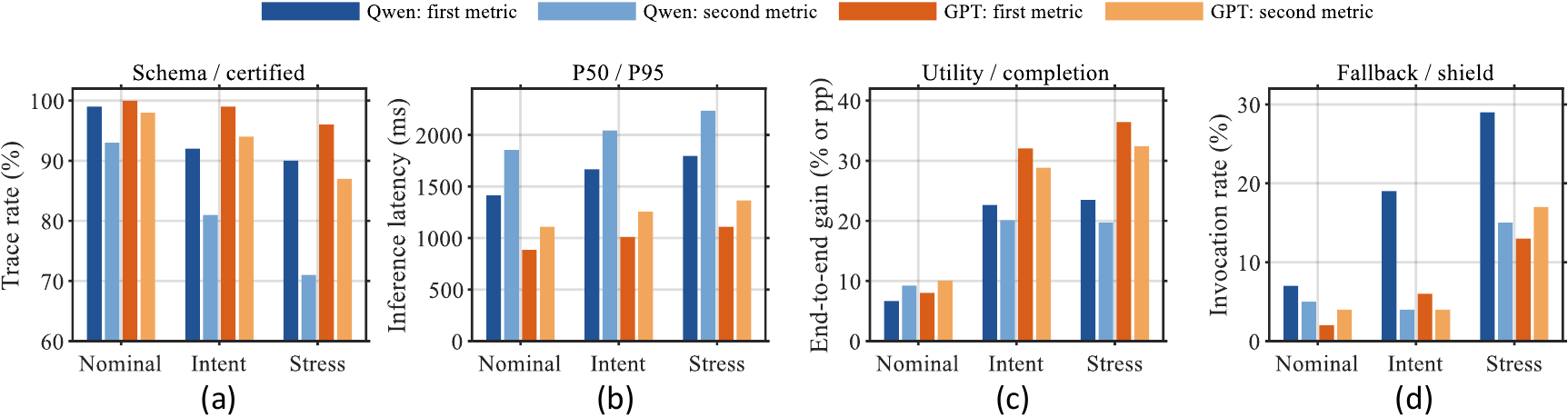}
  \caption{End-to-end calibrated inference-trace replay. For each model, darker/lighter bars represent the first/second metric:
  (a) schema and certification rates;
  (b) P50/P95 inference latency;
  (c) realized utility/completion gain over deterministic-only candidates;
  (d) forced fallback and shield-intervention rates.}
  \label{fig:llm-e2e}
\end{figure*}

The 600-record replay processes visible agent outputs through parsing, certification, joint selection with deterministic fallbacks, execution accounting, and feedback.
Fig.~\ref{fig:llm-e2e} (a) shows declining schema-valid/certified-call rates from nominal to compound stress: 99/93\% to 90/71\% for Qwen and 100/98\% to 96/87\% for GPT. In Fig.~\ref{fig:llm-e2e} (b), both 50th-percentile (P50)
and 95th-percentile (P95) latency increase; Qwen/GPT P95 rises from 1.85/1.11 to 2.24/1.37~s. Fig.~\ref{fig:llm-e2e} (c) shows larger gains under intent shift and compound stress than under nominal conditions. Under stress, the mixed candidate set improves utility by 23.5/36.5\% and completion by 19.7/32.4 percentage points over deterministic-only candidates. Fig.~\ref{fig:llm-e2e} (d) reports stress-state forced-fallback rates of 29/13\% and shield-intervention rates of 15/17\%, with no executed reserve violation.

These results suggest that harder states increase the value of intent-aware proposals while reducing their feasibility. Realized gains therefore depend on both candidate quality and certification success. Forced fallback handles calls without a usable agent candidate, whereas shield intervention records candidate rejection and can coexist with successful selection. The second-scale inference profiles also support keeping orchestration outside the millisecond scheduling path.

\subsection{Multi-User Physical-Layer Validation}
\begin{figure*}[t]
  \centering
  \includegraphics[width=0.95\textwidth]{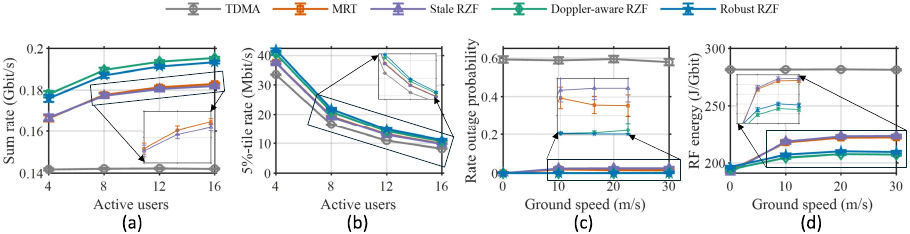}
  \caption{MU-MIMO validation with mean and 95\% confidence interval:
  (a) sum rate and (b) 5th-percentile rate versus active users at 20~m/s;
  (c) 12-Mbit/s outage probability and
  (d) RF energy per delivered Gbit versus speed for 12 users.}
  \label{fig:radio}
       \vspace{-0.6cm}
\end{figure*}

We compare Time-Division Multiple Access (TDMA),
Maximum-Ratio Transmission (MRT), stale RZF,
Doppler-aware RZF, and robust RZF. We use 240 independent geometry/fading trials per point, evaluating actual SINR with all cross-beam interference, while robust design uses \eqref{eq:robustgain}. In Fig.~\ref{fig:radio} (a), sum rate increases with diminishing increments as users are added; robust RZF incurs a 1.1\% sum-rate cost relative to Doppler-aware RZF at 16 users. Fig.~\ref{fig:radio} (b) shows declining cell-edge rates, but robust RZF improves the 16-user 5th-percentile rate from stale RZF's 9.75 to 11.14~Mbit/s. For 12 users, Fig.~\ref{fig:radio} (c) shows stale-RZF outage reaching 2.47\% at 30~m/s, whereas robust RZF records no outage at that speed. Fig.~\ref{fig:radio} (d) shows that Doppler-aware designs limit the mobility-related energy penalty: robust RZF uses 209.4 versus stale RZF's 223.2~J/Gbit at 30~m/s.

The opposing sum-rate and cell-edge-rate trends reflect spatial multiplexing competing with power sharing and residual interference. Doppler compensation and conservative gain margins help limit leakage under aged CSI, preserving delivered rate per joule at a small aggregate-rate cost. These gains arise with explicit multi-user interference included in the evaluation.

\subsection{Load Scaling}

\begin{figure*}[t]
  \centering
  \includegraphics[width=0.95\textwidth]{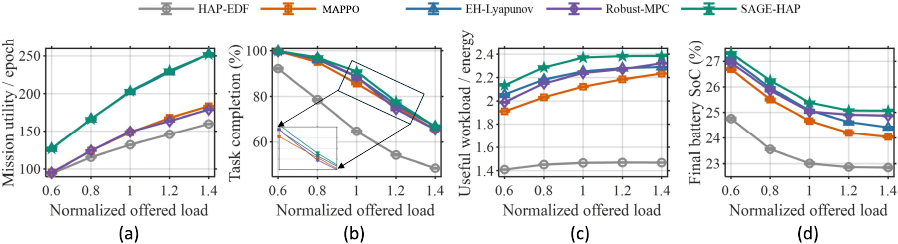}
  \caption{Load scaling with mean and 95\% confidence interval:
  (a) intent-weighted mission utility;
  (b) task completion;
  (c) useful-workload energy efficiency;
  (d) final battery SoC.}
  \label{fig:scaling}
\end{figure*}

Across normalized loads 0.6--1.4, Fig.~\ref{fig:scaling} (a) shows that \SAGE and EH-Lyapunov achieve similar mission utility above the other baselines; their nominal-load values are 203.36 and 202.61. In Fig.~\ref{fig:scaling} (b), completion declines with load, but \SAGE retains 90.94\% at nominal load versus 88.56\% for EH-Lyapunov. Fig.~\ref{fig:scaling} (c) shows increasing, then saturating, useful-workload energy efficiency; the nominal-load values are 2.371 and 2.253, respectively, a 5.27\% gain. Fig.~\ref{fig:scaling} (d) shows declining final SoC as load grows. At load 1.4, \SAGE ends at 25.07\%, whereas EH-Lyapunov reaches 24.42\%, below the 25\% reserve, despite their utility differing by less than 0.1\%.

Higher load offers more valuable work, but finite resource budgets reduce the completed fraction. Similar weighted utility need not imply equal completion: fairness debt protects valuable tasks in both schemes, while adaptive modes can reduce resource costs and accommodate more work. The lower utility of MAPPO and Robust-MPC is consistent with their fixed evaluation objectives under changed mission priorities. HAP-EDF's raw-mode service incurs higher payload costs, helping explain its low efficiency. The SoC contrast further shows why long-term energy balancing requires a separate mechanism for physical reserve protection.

\subsection{Compound-Disturbance Response}
This experiment evaluates how \SAGE and the baselines respond to overlapping traffic surges and declining solar supply. It focuses on critical-service continuity, battery-reserve protection, and delay recovery.

\begin{figure*}[t]
  \centering
  \includegraphics[width=0.95\textwidth]{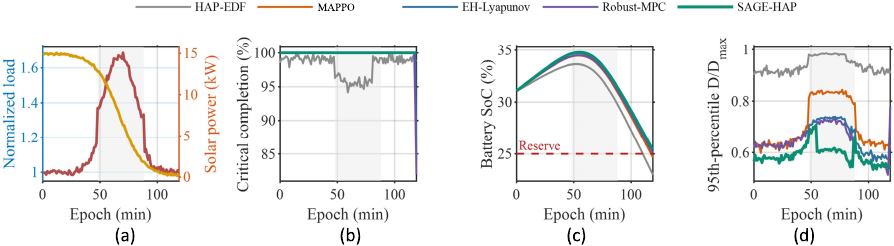}
  \caption{Compound-disturbance response:
  (a) offered load and solar power;
  (b) critical-task completion;
  (c) battery SoC;
  (d) 95th-percentile deadline-normalized delay.
  Shading marks the disturbance interval.}
  \label{fig:dynamic}
       \vspace{-0.6cm}
\end{figure*}

Fig.~\ref{fig:dynamic} (a) shows the two environmental inputs shared by all schemes: normalized offered load on the left axis and solar power on the right axis. The load peaks near epoch 67 while solar input declines, creating concurrent pressure on service capacity and available energy. In Fig.~\ref{fig:dynamic} (b), \SAGE maintains nearly 100\% critical completion, whereas HAP-EDF drops to approximately 95\% during the surge; Robust-MPC also exhibits a final-epoch drop. Fig.~\ref{fig:dynamic} (c) shows initial charging followed by discharging: \SAGE finishes at 25.39\% SoC, while HAP-EDF crosses below the reserve. In Fig.~\ref{fig:dynamic} (d), \SAGE's normalized tail delay briefly approaches 0.70 before returning to about 0.60 during the disturbance, while MAPPO exceeds 0.8 and HAP-EDF has the highest delay.

Discharging begins when solar input no longer covers platform and payload loads. The service response is consistent with increased critical-task priority and semantic-mode switching releasing capacity from monitoring traffic. The lower solar envelope can trigger conservation before the reserve becomes imminent, while feedback-guided replanning may aid delay recovery. MAPPO still adapts actions to observed states, but its frozen reward specification may limit alignment with the changed mission. The terminal MPC completion drop coincides with low battery headroom, suggesting pressure on feasible service near the horizon's end.

\subsection{Forecast Robustness and Reserve Safety}
This sensitivity study compares HAP-EDF, Robust-MPC, and \SAGE as representatives of capacity-driven admission, point-forecast planning, and calibrated reserve certification, respectively. The other two baselines do not rely on solar energy predictions, so they are not included in this comparison.

\begin{figure}[t]
  \centering
  \includegraphics[width=0.47\textwidth]{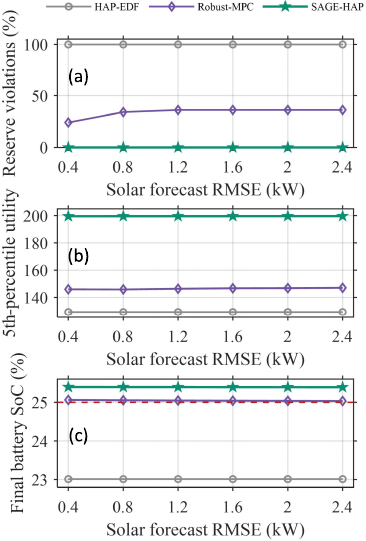}
  \caption{Robustness to solar forecast error:
  (a) fraction of runs with any reserve violation;
  (b) 5th-percentile mission utility;
  (c) final battery SoC.}
  \label{fig:robustness}
       \vspace{-0.6cm}
\end{figure}

In Fig.~\ref{fig:robustness} (a), \SAGE records 0/50 runs with any reserve violation at each tested forecast RMSE from 0.4 to 2.4~kW. Robust-MPC rises from 24\% at 0.4~kW to 36\% at 1.2~kW and above, while HAP-EDF violates the reserve in every run. Fig.~\ref{fig:robustness} (b) shows \SAGE's 5th-percentile utility remaining near 199, compared with 146--147 for Robust-MPC. Fig.~\ref{fig:robustness} (c) places \SAGE's final SoC near 25.39\%, with Robust-MPC close to the 25\% boundary and HAP-EDF below it. An endpoint SoC near the reserve does not exclude earlier violations or variation across runs.

These trends are consistent with lower-envelope certification limiting forecast-driven overcommitment, while semantic adaptation preserves useful service under conservative energy budgets. Channel robustness alone cannot compensate for solar point-forecast errors. The zero violation count is empirical: Proposition~\ref{prop:safety} requires calibrated coverage and bounded actuation error.

\subsection{Ablation Study}

Table~\ref{tab:ablation} compares four variants with full
\SAGE at load 1.2 and 2.0~kW forecast RMSE over 50 runs.
\emph{w/o Intent Translation} removes intent-dependent
priority updates; \emph{w/o Adaptive Modes} uses a fixed semantic-mode map; \emph{w/o Reflection} disables feedback-driven plan revision; and \emph{w/o Energy Certificate} bypasses battery-reserve
certification. All variants are evaluated using the active
mission weights.

\begin{table*}[t!]
\caption{Ablation Results for SAGE-HAP Under Stressed Load and Solar Uncertainty.}
\label{tab:ablation}
\renewcommand{\arraystretch}{1.18}
\renewcommand{\arrayrulewidth}{0.8pt}
\renewcommand{\tabcolsep}{2.5pt}
{\fontsize{7.1}{8.6}\selectfont
\makebox[\linewidth][c]{%
\begin{tabular}{lccccccc}
\hline\hline
\rowcolor{graymid}
\textbf{Configuration}
& \shortstack{\textbf{Utility}$\uparrow$}
& \shortstack{\textbf{Completion}\textbf{(\%) $\uparrow$}}
& \shortstack{\textbf{Critical Comp.}\textbf{(\%) $\uparrow$}}
& \shortstack{\textbf{Useful Work/}\textbf{Energy $\uparrow$}}
& \shortstack{\textbf{P95 Delay}\textbf{(ms) $\downarrow$}}
& \shortstack{\textbf{Final SoC}\textbf{(\%) $\uparrow$}}
& \shortstack{\textbf{Violating Runs}\textbf{(\%) $\downarrow$}} \\
\hline\hline

\rowcolor{greenlight}
w/o Intent Translation
& $171.04 \pm 1.53$
& $81.97 \pm 0.85$
& $99.61 \pm 0.10$
& $2.406 \pm 0.010$
& $932.02 \pm 11.9$
& $25.18 \pm 0.02$
& $4.0~(2/50)$ \\

w/o Adaptive Modes
& $227.66 \pm 1.34$
& $74.28 \pm 1.02$
& $99.84 \pm 0.04$
& $2.315 \pm 0.014$
& $936.31 \pm 13.4$
& $25.08 \pm 0.02$
& $8.0~(4/50)$ \\

\rowcolor{greenlight}
w/o Reflection
& $231.24 \pm 1.42$
& $79.39 \pm 0.94$
& $99.90 \pm 0.02$
& $2.392 \pm 0.011$
& $954.82 \pm 16.2$
& $25.06 \pm 0.01$
& $10.0~(5/50)$ \\

w/o Energy Certificate
& $235.32 \pm 1.76$
& $81.73 \pm 1.08$
& $99.98 \pm 0.01$
& $2.412 \pm 0.013$
& $926.59 \pm 12.5$
& $24.81 \pm 0.03$
& $94.0~(47/50)$ \\

\rowcolor{greenstrong}
\textbf{SAGE-HAP (Full)}
& \textbf{$230.62 \pm 1.22$}
& \textbf{$77.89 \pm 0.82$}
& \textbf{$99.91 \pm 0.02$}
& \textbf{$2.384 \pm 0.009$}
& \textbf{$923.79 \pm 11.1$}
& \textbf{$25.12 \pm 0.01$}
& \textbf{$2.0~(1/50)$} \\
\hline\hline
\end{tabular}%
}
}
\end{table*}

Removing intent translation increases unweighted completion but reduces mission utility by 25.8\%.
Without adaptive modes, completion and energy efficiency
decrease by 3.61 percentage points and 2.89\%, respectively.
Without reflection, P95 delay increases by 3.36\%, while
utility increases by only 0.27\%.
Removing the energy certificate raises utility by 2.04\%
and completion by 3.84 percentage points, but lowers final
SoC to 24.81\% and produces reserve violations in 47/50 runs.
The full method finishes at 25.12\% SoC with 1/50 violating runs.

These patterns suggest that intent translation prevents
numerous low-value tasks from displacing work prioritized
by the emergency mission, explaining why higher completion
can accompany lower utility. Adaptive semantic modes reduce transmission and computation requirements, allowing more tasks to fit within the available resource budget. Reflection primarily improves service timing through feedback-based corrections, accounting for the delay benefit despite little change in utility. Energy certification limits commitments that would compromise the protected reserve, trading some immediate service for sustained operation. The full method's nonzero empirical violation rate remains consistent with a safety guarantee conditional on solar coverage and bounded actuation error.

\section{Conclusion}
\label{sec:conclusion}

This paper presented \SAGE, a sustainability-certified agentic AI framework for a solar-powered HAP communication--computing anchor. The agent perceives state, translates intent, diagnoses bottlenecks, and proposes typed admission, mode, and grouping plans; deterministic RZF/refinement and an energy shield alone authorize execution. 

The 600-record end-to-end replay shows how schema errors and certification failures activate safe fallbacks: under compound stress, Qwen/GPT profiles retain certified agent candidates in 71/87\% of calls and provide 23.5/36.5\% realized utility gains without a shielded reserve violation. MU-MIMO tests further confirm that robust RZF trades a small sum-rate loss for higher cell-edge reliability. Finally, removing the energy certificate violates the reserve in 94\% of stressed runs. Thus, \SAGE makes agentic adaptability useful only after its radio, computing, and energy consequences become certifiable.

\bibliographystyle{IEEEtran}
\bibliography{sage_hap_references}

@article{kurt2021vision,
  author  = {G. K. Kurt and M. G. Khoshkholgh and S. Alfattani and A. Ibrahim and T. S. J. Darwish and M. S. Alam and H. Yanikomeroglu and A. Yongacoglu},
  title   = {A Vision and Framework for the High Altitude Platform Station Networks of the Future},
  journal = {IEEE Communications Surveys \& Tutorials},
  volume  = {23},
  number  = {2},
  pages   = {729--779},
  year    = {2021},
  doi     = {10.1109/COMST.2021.3066905}
}

@article{arum2020review,
  author  = {S. C. Arum and D. Grace and P. D. Mitchell},
  title   = {A Review of Wireless Communication Using High-Altitude Platforms for Extended Coverage and Capacity},
  journal = {Computer Communications},
  volume  = {157},
  pages   = {232--256},
  year    = {2020},
  doi     = {10.1016/j.comcom.2020.04.020}
}

@article{arum2024twotier,
  author  = {S. C. Arum and H. Yanikomeroglu},
  title   = {Extending Coverage and Capacity From High Altitude Platforms With a Two-Tier Cellular Architecture},
  journal = {IEEE Transactions on Mobile Computing},
  volume  = {23},
  number  = {2},
  pages   = {1942--1953},
  year    = {2024}
}

@article{javed2025system,
  author  = {S. Javed and M.-S. Alouini},
  title   = {System Design and Parameter Optimization for Remote Coverage From {NOMA}-Based High-Altitude Platform Stations},
  journal = {IEEE Transactions on Wireless Communications},
  volume  = {24},
  number  = {2},
  pages   = {1387--1400},
  year    = {2025},
  doi     = {10.1109/TWC.2024.3508872}
}

@inproceedings{lakew2021offloading,
  author    = {D. S. Lakew and A.-T. Tran and N.-N. Dao and S. Cho},
  title     = {Intelligent Offloading and Resource Allocation in {HAP}-Assisted {MEC} Networks},
  booktitle = {Proc. International Conference on Information and Communication Technology Convergence (ICTC)},
  pages     = {1582--1587},
  year      = {2021}
}

@article{traspadini2023realtime,
  author  = {A. Traspadini and M. Giordani and G. Giambene and M. Zorzi},
  title   = {Real-Time {HAP}-Assisted Vehicular Edge Computing for Rural Areas},
  journal = {IEEE Wireless Communications Letters},
  volume  = {12},
  number  = {4},
  pages   = {674--678},
  year    = {2023},
  doi     = {10.1109/LWC.2023.3238851}
}

@article{ren2023handoff,
  author  = {Q. Ren and O. Abbasi and G. K. Kurt and H. Yanikomeroglu and J. Chen},
  title   = {Handoff-Aware Distributed Computing in High Altitude Platform Station-Assisted Vehicular Networks},
  journal = {IEEE Transactions on Wireless Communications},
  volume  = {22},
  number  = {12},
  pages   = {8814--8827},
  year    = {2023}
}

@article{abderrahim2023green,
  author  = {W. Abderrahim and O. Amin and B. Shihada},
  title   = {How to Leverage High Altitude Platforms in Green Computing?},
  journal = {IEEE Communications Magazine},
  volume  = {61},
  number  = {7},
  pages   = {134--140},
  year    = {2023},
  doi     = {10.1109/MCOM.001.2200732}
}

@article{abderrahim2024datacenter,
  author  = {W. Abderrahim and O. Amin and B. Shihada},
  title   = {Data Center-Enabled High Altitude Platforms: A Green Computing Alternative},
  journal = {IEEE Transactions on Mobile Computing},
  volume  = {23},
  number  = {5},
  pages   = {6149--6162},
  year    = {2024},
  doi     = {10.1109/TMC.2023.3316204}
}

@article{javed2023interdisciplinary,
  author  = {S. Javed and M.-S. Alouini and Z. Ding},
  title   = {An Interdisciplinary Approach to Optimal Communication and Flight Operation of High-Altitude Long-Endurance Platforms},
  journal = {IEEE Transactions on Aerospace and Electronic Systems},
  volume  = {59},
  number  = {6},
  pages   = {8327--8341},
  year    = {2023},
  doi     = {10.1109/TAES.2023.3304615}
}

@article{alhourani2014pathloss,
  author  = {A. Al-Hourani and S. Kandeepan and S. Lardner},
  title   = {Optimal {LAP} Altitude for Maximum Coverage},
  journal = {IEEE Wireless Communications Letters},
  volume  = {3},
  number  = {6},
  pages   = {569--572},
  year    = {2014},
  doi     = {10.1109/LWC.2014.2342736}
}

@article{xie2021deep,
  author  = {H. Xie and Z. Qin and G. Y. Li and B.-H. Juang},
  title   = {Deep Learning Enabled Semantic Communication Systems},
  journal = {IEEE Transactions on Signal Processing},
  volume  = {69},
  pages   = {2663--2675},
  year    = {2021},
  doi     = {10.1109/TSP.2021.3071210}
}

@article{strinati2021semantic,
  author  = {E. C. Strinati and S. Barbarossa},
  title   = {6G Networks: Beyond Shannon Towards Semantic and Goal-Oriented Communications},
  journal = {Computer Networks},
  volume  = {190},
  pages   = {107930},
  year    = {2021},
  doi     = {10.1016/j.comnet.2021.107930}
}

@inproceedings{cang2023semantic,
  author    = {Y. Cang and M. Chen and Z. Yang and Y. Hu and Y. Wang and Z. Zhang and K.-K. Wong},
  title     = {Resource Allocation for Semantic-Aware Mobile Edge Computing Systems},
  booktitle = {Proc. IEEE Global Communications Conference Workshops},
  pages     = {1585--1590},
  year      = {2023}
}

@article{nguyen2026semantic,
  author  = {L. X. Nguyen and others},
  title   = {A Contemporary Survey on Semantic Communications: Theory of Mind, Generative {AI}, and Deep Joint Source-Channel Coding},
  journal = {IEEE Communications Surveys \& Tutorials},
  volume  = {28},
  pages   = {2377--2417},
  year    = {2026}
}

@book{neely2010stochastic,
  author    = {M. J. Neely},
  title     = {Stochastic Network Optimization with Application to Communication and Queueing Systems},
  publisher = {Morgan \& Claypool},
  address   = {San Rafael, CA, USA},
  year      = {2010}
}

@article{huang2013utility,
  author  = {L. Huang and M. J. Neely},
  title   = {Utility Optimal Scheduling in Energy-Harvesting Networks},
  journal = {IEEE/ACM Transactions on Networking},
  volume  = {21},
  number  = {4},
  pages   = {1117--1130},
  year    = {2013},
  doi     = {10.1109/TNET.2012.2222195}
}

@article{ozel2012awgn,
  author  = {O. Ozel and S. Ulukus},
  title   = {Achieving {AWGN} Capacity Under Stochastic Energy Harvesting},
  journal = {IEEE Transactions on Information Theory},
  volume  = {58},
  number  = {10},
  pages   = {6471--6483},
  year    = {2012},
  doi     = {10.1109/TIT.2012.2208175}
}

@article{angelopoulos2023conformal,
  author  = {A. N. Angelopoulos and S. Bates},
  title   = {Conformal Prediction: A Gentle Introduction},
  journal = {Foundations and Trends in Machine Learning},
  volume  = {16},
  number  = {4},
  pages   = {494--591},
  year    = {2023},
  doi     = {10.1561/2200000101}
}

@book{boyd2004convex,
  author    = {S. Boyd and L. Vandenberghe},
  title     = {Convex Optimization},
  publisher = {Cambridge University Press},
  address   = {Cambridge, U.K.},
  year      = {2004}
}

@article{wang2024agents,
  author  = {L. Wang and C. Ma and X. Feng and Z. Zhang and H. Yang and J. Zhang and Z. Chen and J. Tang and X. Chen and Y. Lin and W. X. Zhao and Z. Wei and J.-R. Wen},
  title   = {A Survey on Large Language Model Based Autonomous Agents},
  journal = {Frontiers of Computer Science},
  volume  = {18},
  number  = {6},
  pages   = {186345},
  year    = {2024},
  doi     = {10.1007/s11704-024-40231-1}
}

@article{sarhan2026beamforming,
  author  = {A. Y. Sarhan and O. A. Abdullah and K. T. Musri and H. Al-Hraishawi},
  title   = {Physics-Informed Uncertainty-Aware Beamforming for {HAPS} Massive {MIMO} under Imperfect {CSI}},
  journal = {arXiv preprint arXiv:2606.29074},
  year    = {2026}
}

@article{peel2005rzf,
  author  = {C. B. Peel and B. M. Hochwald and A. L. Swindlehurst},
  title   = {A Vector-Perturbation Technique for Near-Capacity Multiantenna Multiuser Communication---Part {I}: Channel Inversion and Regularization},
  journal = {IEEE Transactions on Communications},
  volume  = {53},
  number  = {1},
  pages   = {195--202},
  year    = {2005},
  doi     = {10.1109/TCOMM.2004.840638}
}

@article{liu2026tools,
  title={Unleashing Tool Engineering and Intelligence for Agentic AI in Next-Generation Communication Networks},
  author={Liu, Yinqiu and Zhang, Ruichen and Niyato, Dusit and Jamalipour, Abbas and Duong, Trung Q and Kim, Dong In},
  journal={IEEE Communications Magazine},
  year={2026},
  publisher={IEEE}
}

@article{akhtarshenas2026ppo,
  author  = {Akhtarshenas, Azim and Svistunov, German
             and Bernab{\`e}, Matteo and Zheng, Kuangyu
             and L{\'o}pez-P{\'e}rez, David},
  title   = {{PPO}-Based Dynamic Positioning of {HAPS-BS}
             in Wind-Disturbed Stratospheric Maritime Networks},
  journal = {arXiv preprint arXiv:2605.05240},
  year    = {2026},
}

@article{nkurunziza2026attention,
  title={Attention--based HAPS--to--ground nodes optimization for differential privacy towards secure semantic communications},
  author={Nkurunziza, Pascal and Umehara, Daisuke},
  journal={Neural Computing and Applications},
  volume={38},
  number={5},
  pages={105},
  year={2026},
  publisher={Springer}
}

@article{liang2025knowledge,
  title={Knowledge graph fusion based semantic communication framework},
  author={Liang, Chengsi and Sun, Yao and Niyato, Dusit and Imran, Muhammad Ali},
  journal={IEEE Transactions on Mobile Computing},
  volume={24},
  number={11},
  pages={11416--11429},
  year={2025},
  publisher={IEEE}
}

@inproceedings{minani2024qosem,
  title={QoSem-based resource allocation for semantic communication in NTN downlinks},
  author={Minani, Frodouard and Kobayashi, Makoto and Fujihashi, Takuya and Alim, Md Abdul and Saruwatari, Shunsuke and Watanabe, Takashi},
  booktitle={2024 IEEE Global Communications Conference},
  pages={2587--2592},
  year={2024},
  organization={IEEE}
}

@article{huang2026joint1,
  title={Joint 3D Trajectory and Power Allocation for HAPs--UAV Bistatic ISARAC in Low-Altitude Networks},
  author={Huang, Bang and Alouini, Mohamed-Slim},
  journal={IEEE Internet of Things Journal},
  year={2026},
  publisher={IEEE}
}

@article{huang2026joint,
  title={Joint trajectory and resource optimization for HAPs-SAR systems with energy-aware constraints},
  author={Huang, Bang and Park, Ki-Hong and Pang, Xiaowei and Alouini, Mohamed-Slim},
  journal={IEEE Transactions on Wireless Communications},
  volume={25},
  pages={21134--21148},
  year={2026},
  publisher={IEEE}
}

@article{zhang2025maintaining,
  title={Maintaining Predictable QoS for Online Service Provisioning in Non-Terrestrial Networks via Safe Transfer Learning},
  author={Zhang, Shengyu and Dou, Songshi and Li, Zhenglong and Yeung, Kwan L and Quek, Tony QS},
  journal={IEEE Transactions on Networking},
  volume={34},
  pages={1745--1760},
  year={2025},
  publisher={IEEE}
}

@article{wang2026lyapunov,
  title={Lyapunov-Based Tri-Stage Online On-Demand Resource Allocation and Task Offloading in SAGIN},
  author={Wang, Luqiao and Li, Changle and Zhang, Yao and Yue, Wenwei and Sha, Zifan and Mashhadi, Mahdi Boloursaz and Sun, Zhili and Cheng, Nan and Yu, F Richard},
  journal={IEEE Transactions on Wireless Communications},
  year={2026},
  publisher={IEEE}
}

@article{minani2025hybrid,
  title={Hybrid Semantic-Bit Communications for NTN Downlinks: A Hierarchical Multi-Agent DRL Approach},
  author={Minani, Frodouard and Kobayashi, Makoto and Fujihashi, Takuya and Alim, MD Abdul and Saruwatari, Shunsuke and Nishi, Masahiro and Watanabe, Takashi},
  journal={IEEE Open Journal of the Communications Society},
  year={2025},
  publisher={IEEE}
}

@article{nkurunziza2025deepstas,
  title={DeepSTAS: dl-assisted semantic transmission accuracy enhancement through an attention-driven HAPS relay system},
  author={Nkurunziza, Pascal and Umehara, Daisuke},
  journal={Technologies},
  volume={13},
  number={4},
  pages={137},
  year={2025},
  publisher={MDPI}
}

@article{luo2026trustworthy,
  title={A trustworthy agentic multi-llm network: Challenges, solutions, and a use case},
  author={Luo, Haoxiang and Sun, Gang and Liu, Yinqiu and Niyato, Dusit and Yu, Hongfang and Atiquzzaman, Mohammed and Dustdar, Schahram},
  journal={IEEE Wireless Communications},
  year={2026},
  publisher={IEEE}
}

@article{li2026efficient,
  author  = {Li, Wanyue and Li, Shuyang and Hao, Jie
             and Wu, Qiang and Wang, Ran},
  title   = {Efficient Task Offloading and Resource Allocation
             in {HAPS}-Assisted {LEO} Satellite Networks:
             A {MAPPO} With Exact Potential Game Approach},
  journal = {IEEE Internet of Things Journal},
  volume  = {13},
  number  = {6},
  pages   = {10179--10195},
  year    = {2026},
}

@article{wang2025toward,
  title={Toward intelligent space-air-ground integrated network: Architecture, challenges, and emerging directions},
  author={Wang, Lina and Fan, Mingrui and Yang, Ning and Ma, Xu and Liang, Yan and Zhang, Haijun},
  journal={Journal of Communications and Information Networks},
  volume={10},
  number={2},
  pages={87--102},
  year={2025},
  publisher={PTP}
}

@article{luo2026best,
  title={BEST: Blockchain-Enabled Sustainable Task Scheduling for Optimizing Large AI Model Workloads at the Edge},
  author={Luo, Haoxiang and Chen, Runhua and Sun, Gang and Yu, Hongfang and Dustdar, Schahram},
  journal={IEEE Transactions on Artificial Intelligence},
  year={2026},
  publisher={IEEE}
}

@article{yan2026hierarchical,
  title={Hierarchical LLM-Driven Control for HAPS-Assisted UAV Networks: Joint Optimization of Flight and Connectivity},
  author={Yan, Zijiang and Zhou, Hao and Jaafar, Wael and Pei, Jianhua and Wang, Ping and Yanikomeroglu, Halim and Tabassum, Hina},
  journal={arXiv preprint arXiv:2605.11509},
  year={2026}
}

@article{luo2024communication,
  title={Communication-efficient federated learning with adaptive aggregation for heterogeneous client-edge-cloud network},
  author={Luo, Long and Zhang, Chi and Yu, Hongfang and Sun, Gang and Luo, Shouxi and Dustdar, Schahram},
  journal={IEEE Transactions on Services Computing},
  volume={17},
  number={6},
  pages={3241--3255},
  year={2024},
  publisher={IEEE}
}

@article{luo2025convergence,
  title={Convergence of symbiotic communications and blockchain for sustainable and trustworthy 6G wireless networks},
  author={Luo, Haoxiang and Sun, Gang and Chi, Cheng and Yu, Hongfang and Guizani, Mohsen},
  journal={IEEE Wireless Communications},
  volume={32},
  number={2},
  pages={18--25},
  year={2025},
  publisher={IEEE}
}

@inproceedings{luo2026agentic,
  title={Agentic AI-Enabled Space-Air Integrated Computing Power Network (SAICPN) for Efficient Task Execution in 6G},
  author={Luo, Haoxiang and Zheng, Runze and Zhang, Ruichen and Liu, Yinqiu and Sun, Gang and Yu, Hongfang and Guizani, Mohsen},
  booktitle={2026 International Wireless Communications and Mobile Computing (IWCMC)},
  pages={827--832},
  year={2026},
  organization={IEEE}
}

@article{luo2026toward,
  title={High-Altitude Platforms Beyond Connectivity: A Survey of Integrated Sensing, Storage, Communication, Computing, and Intelligence},
  author={Luo, Haoxiang and Alouini, Mohamed-Slim},
  journal={arXiv preprint arXiv:2608.18587},
  year={2026}
}

@article{shokouhi2026intent,
  author  = {M. H. Shokouhi and V. W. S. Wong},
  title   = {Agentic {AI} for Intent-Driven Optimization in Cell-Free {O-RAN}},
  journal = {arXiv preprint arXiv:2602.22539},
  year    = {2026}
}

@inproceedings{han2025sky,
  author    = {D. Han and A. Zhang and R. Chen and C. Feng and S. Guo},
  title     = {Agent in the Sky: Intelligent Multi-Agent Framework for Autonomous {HAPS} Coordination and Real-World Event Adaptation},
  booktitle = {AAAI Workshop on Artificial Intelligence for Wireless Communications and Networking},
  pages     = {1--6},
  year      = {2025}
}

@article{luo2026background,
  title={Agentic AI-Enabled Solar-Powered High-Altitude Platforms for Sustainable SAGINs},
  author={Luo, Haoxiang and Huang, Bang and Alouini, Mohamed-Slim},
  journal={arXiv preprint arXiv:2608.15087},
  year={2026}
}

@techreport{3gpp38811,
  author      = {{3GPP}},
  title       = {Study on New Radio to Support Non-Terrestrial Networks},
  institution = {3rd Generation Partnership Project},
  number      = {TR 38.811, Version 15.4.0},
  year        = {2020}
}

@article{schmalstieg2014aging,
  author  = {J. Schmalstieg and S. K{"a}bitz and M. Ecker and D. U. Sauer},
  title   = {A Holistic Aging Model for {Li(NiMnCo)O2} Based 18650 Lithium-Ion Batteries},
  journal = {Journal of Power Sources},
  volume  = {257},
  pages   = {325--334},
  year    = {2014},
  doi     = {10.1016/j.jpowsour.2014.02.012}
}

\end{document}